\input harvmac
\input epsf

\input amssym.def
\input epsf

\noblackbox
\newcount\figno
\figno=0
\def\fig#1#2#3{
\par\begingroup\parindent=0pt\leftskip=1cm\rightskip=1cm\parindent=0pt
\baselineskip=11pt
\global\advance\figno by 1
\midinsert
\epsfxsize=#3
\centerline{\epsfbox{#2}}
\vskip 12pt
\centerline{{\bf Figure \the\figno} #1}\par
\endinsert\endgroup\par}
\def\figlabel#1{\xdef#1{\the\figno}}

\def\pano{\par\noindent}

\def\pmb#1{\setbox0=\hbox{#1}%
 \kern-.025em\copy0\kern-\wd0
 \kern.05em\copy0\kern-\wd0
 \kern-.025em\raise.0433em\box0 }
\font\cmss=cmss10
\font\cmsss=cmss10 at 7pt
\def\rlx{\relax\leavevmode}
\def\Cop{\relax\,\hbox{$\inbar\kern-.3em{\rm C}$}}
\def\Rop{\relax{\rm I\kern-.18em R}}
\def\Nop{\relax{\rm I\kern-.18em N}}
\def\Pop{\relax{\rm I\kern-.18em P}}
\def\Zop{\rlx\leavevmode\ifmmode\mathchoice{\hbox{\cmss Z\kern-.4em Z}}
 {\hbox{\cmss Z\kern-.4em Z}}{\lower.9pt\hbox{\cmsss Z\kern-.36em Z}}
 {\lower1.2pt\hbox{\cmsss Z\kern-.36em Z}}\else{\cmss Z\kern-.4em
 Z}\fi}


\def\AdSs5{$AdS_5$}
\def\AdS5s5{$AdS_5 \times S^5$}



\def\eg{{\it e.g.}}
\def\alphatilde{\tilde{\alpha}}

\def\IZ{\relax\ifmmode\mathchoice {\hbox{\cmss Z\kern-.4em
Z}}{\hbox{\cmss Z\kern-.4em Z}} {\lower.9pt\hbox{\cmsss Z\kern-.4em
Z}} {\lower1.2pt\hbox{\cmsss Z\kern-.4em Z}}\else{\cmss Z\kern-.4em
Z}\fi}

\def\c1{{\chi^1}}

\def\N4{{\cal N}=4}
\def\half{{1\over 2}}

\parindent 25pt
\overfullrule=0pt
\tolerance=10000
\def\ie{{\it i.e.}}


\def\pmb#1{\setbox0=\hbox{#1}%
 \kern-.025em\copy0\kern-\wd0
 \kern.05em\copy0\kern-\wd0
 \kern-.025em\raise.0433em\box0 }
\font\cmss=cmss10
\font\cmsss=cmss10 at 7pt
\def\rlx{\relax\leavevmode}
\def\Zop{\rlx\leavevmode\ifmmode\mathchoice{\hbox{\cmss Z\kern-.4em Z}}
 {\hbox{\cmss Z\kern-.4em Z}}{\lower.9pt\hbox{\cmsss Z\kern-.36em Z}}
 {\lower1.2pt\hbox{\cmsss Z\kern-.36em Z}}\else{\cmss Z\kern-.4em
 Z}\fi}


\def\pmb#1{\setbox0=\hbox{#1}%
 \kern-.025em\copy0\kern-\wd0
 \kern.05em\copy0\kern-\wd0
 \kern-.025em\raise.0433em\box0 }
\font\cmss=cmss10
\font\cmsss=cmss10 at 7pt
\def\rlx{\relax\leavevmode}
\def\inbar{\,\vrule height1.5ex width.4pt depth0pt}
\def\Cop{\relax\,\hbox{$\inbar\kern-.3em{\rm C}$}}
\def\Rop{\relax{\rm I\kern-.18em R}}
\def\Nop{\relax{\rm I\kern-.18em N}}
\def\Pop{\relax{\rm I\kern-.18em P}}

\def\Zop{\rlx\leavevmode\ifmmode\mathchoice{\hbox{\cmss Z\kern-.4em Z}}
 {\hbox{\cmss Z\kern-.4em Z}}{\lower.9pt\hbox{\cmsss Z\kern-.36em Z}}
 {\lower1.2pt\hbox{\cmsss Z\kern-.36em Z}}\else{\cmss Z\kern-.4em
 Z}\fi}

\def\Res{{\Res}}



\lref\metsaev{
R.~R.~Metsaev,
{\it Type IIB Green-Schwarz superstring in plane wave Ramond-Ramond
background},
Nucl.\ Phys.\ B {\bf 625}, 70 (2002); {\tt hep-th/0112044}.
}

\lref\metsaevtseytlin{
R.~R.~Metsaev, A.~A.~Tseytlin,
{\it Exactly solvable model of superstring in plane wave Ramond-Ramond  background,}
Phys.\ Rev.\ D {\bf 65}, 126004 (2002);
{\tt hep-th/0202109}.}

\lref\bmn{
D.~Berenstein, J.~M.~Maldacena, H.~Nastase,
{\it Strings in flat space and pp waves from N = 4 super Yang Mills,}
JHEP {\bf 0204}, 013 (2002);
{\tt hep-th/0202021}.}

\lref\GSstringfield{
M.~B.~Green, J.~H.~Schwarz,
{\it Superstring Field Theory},
Nucl.\ Phys.\ B {\bf 243}, 475 (1984).}

\lref\gstwo{
M.~B.~Green, J.~H.~Schwarz,
{\it Superstring Interactions,}
Nucl.\ Phys.\ B {\bf 218}, 43 (1983).}

\lref\spradlinvolovich{
M.~Spradlin, A.~Volovich,
{\it Superstring interactions in a pp-wave background,}
Phys.\ Rev.\ D {\bf 66}, 086004 (2002);
{\tt hep-th/0204146}.}

\lref\spradlintwo{
M.~Spradlin, A.~Volovich,
{\it Superstring interactions in a pp-wave background II,}
JHEP {\bf 0301}, 036 (2003);
{\tt hep-th/0206073}.}

\lref\klebanovspradlinvolovich{
I.~R.~Klebanov, M.~Spradlin, A.~Volovich,
{\it New Effects in Gauge Theory from pp-wave Superstrings,}
Phys.\ Lett.\ B {\bf 548}, 111 (2002);
{\tt hep-th/0206221}.}

\lref\pankstef{
A.~Pankiewicz, B.~J.~Stefanski,
{\it pp-wave light-cone superstring field theory,}
Nucl.\ Phys.\ B {\bf 657}, 79 (2003);
{\tt hep-th/0210246}.}

\lref\gomis{
J.~Gomis, S.~Moriyama, J.~w.~Park,
{\it Open + closed string field theory from gauge fields;}
{\tt hep-th/0305264}.}

\lref\bogdanone{
B.~J.~Stefanski,
{\it Open string plane-wave light-cone superstring field theory;}
{\tt hep-th/0304114}.}

\lref\chandra{
B.~Chandrasekhar, A.~Kumar,
{\it D-branes in pp-wave light cone string field theory,}
JHEP {\bf 0306}, 001 (2003);
{\tt hep-th/0303223}.}

\lref\bgmnn{
D.~Berenstein, E.~Gava, J.~M.~Maldacena, K.~S.~Narain, H.~Nastase,
{\it Open strings on plane waves and their Yang-Mills duals};
{\tt hep-th/0203249}.}

\lref\he{
Y.~H.~He, J.~H.~Schwarz, M.~Spradlin, A.~Volovich,
{\it Explicit formulas for Neumann coefficients in the plane-wave geometry,}
Phys.\ Rev.\ D {\bf 67}, 086005 (2003);
{\tt hep-th/0211198}.}

\lref\Schwarz{
J.~H.~Schwarz,
{\it Comments on superstring interactions in a plane-wave background},
JHEP {\bf 0209}, 058 (2002);
{\tt hep-th/0208179}.}

\lref\lee{
P.~Lee, J.~w.~Park,
{\it Open strings in PP-wave background from defect conformal field theory,}
Phys.\ Rev.\ D {\bf 67}, 026002 (2003)
{\tt hep-th/0203257}.}

\lref\bgg{
O.~Bergman, M.~R.~Gaberdiel, M.~B.~Green,
{\it D-brane interactions in type IIB plane-wave background},
JHEP {\bf 0303}, 002 (2003);
{\tt hep-th/0205183}.}

\lref\gg{
M.~R.~Gaberdiel, M.~B.~Green,
{\it The D-instanton and other supersymmetric D-branes in IIB plane-wave string theory,}
Annals Phys.\  {\bf 307}, 147 (2003);
{\tt hep-th/0211122}.}

\lref\oblique{
M.~R.~Gaberdiel, M.~B.~Green, S.~Schafer-Nameki, A.~Sinha,
{\it Oblique and curved D-branes in IIB plane-wave string theory,}
JHEP {\bf 0310}, 052 (2003);
{\tt hep-th/0306056}.}

\lref\stone{
K.~Skenderis, M.~Taylor,
{\it Branes in AdS and pp-wave spacetimes},
JHEP {\bf 0206}, 025 (2002);
{\tt hep-th/0204054}.}

\lref\sttwo{
K.~Skenderis, M.~Taylor,
{\it Open strings in the plane wave background. I: Quantization and
symmetries}, 
Nucl.\ Phys.\ B {\bf 665}, 3 (2003);
{\tt hep-th/0211011}.}

\lref\stthree{
K.~Skenderis, M.~Taylor,
{\it Open strings in the plane wave background. II: Superalgebras and
spectra}, 
JHEP {\bf 0307}, 006 (2003);
{\tt hep-th/0212184}.}

\lref\Dabholkar{
A.~Dabholkar, S.~Parvizi,
{\it Dp branes in pp-wave background},
Nucl.\ Phys.\ B {\bf 641}, 223 (2002);
{\tt hep-th/0203231}.}

\lref\Billo{
M.~Billo, I.~Pesando,
{\it Boundary states for GS superstrings in an Hpp wave background},
Phys.\ Lett.\ B {\bf 536}, 121 (2002);
{\tt hep-th/0203028}.}

\lref\nayak{
A.~Kumar, R.~R.~Nayak, Sanjay,
{\it D-brane solutions in pp-wave background},
Phys.\ Lett.\ B {\bf 541}, 183 (2002);
{\tt hep-th/0204025}.}

\lref\schnitzer{
S.~G.~Naculich, H.~J.~Schnitzer, N.~Wyllard,
{\it pp-wave limits and orientifolds,}
Nucl.\ Phys.\ B {\bf 650}, 43 (2003)
{\tt hep-th/0206094}.}

\lref\bala{
V.~Balasubramanian, M.~x.~Huang, T.~S.~Levi, A.~Naqvi,
{\it Open strings from N = 4 super Yang-Mills,}
JHEP {\bf 0208}, 037 (2002)
{\tt hep-th/0204196.}}

\lref\asnvs{
A.~Sinha, N.~V.~Suryanarayana,
{\it Tadpole analysis of orientifolded plane-waves,}
JHEP {\bf 0211}, 026 (2002);
{\tt hep-th/0209247}.}

\lref\GreenWai{
M.~B.~Green, P.~Wai,
{\it The Insertion Of Boundaries In World Sheets},
Nucl.\ Phys.\ B {\bf 431}, 131 (1994);}

\lref\Michelson{
J.~Michelson,
{\it (Twisted) toroidal compactification of pp-waves},
Phys.\ Rev.\ D {\bf 66}, 066002 (2002);
{\tt hep-th/0203140}.}

\lref\Japanese{
S.~Mizoguchi, T.~Mogami, Y.~Satoh,
{\it A note on T-duality of strings in plane-wave backgrounds},
Phys.\ Lett.\ B {\bf 564}, 132 (2003);
{\tt hep-th/0302020}.
}

\lref\ShapiroThorn{
J.~A.~Shapiro and C.~B.~Thorn,
{\it Closed String - Open String Transitions And Witten's String Field Theory},
Phys.\ Lett.\ B {\bf 194}, 43 (1987).}

\lref\ShapiroThornone{
J.~A.~Shapiro, C.~B.~Thorn,
{\it Closed String - Open String Transitions And Witten's String Field Theory},
Phys.\ Lett.\ B {\bf 194}, 43 (1987).}

\lref\ShapiroThorntwo{
J.~A.~Shapiro, C.~B.~Thorn,
{\it BRST Invariant Transitions Between Closed And Open Strings},
Phys.\ Rev.\ D {\bf 36}, 432 (1987).}

\lref\Thorn{
C.~B.~Thorn,
{\it The Theory Of Interacting Relativistic Strings},
Nucl.\ Phys.\ B {\bf 263}, 493 (1986).}

\lref\GilesThorn{
R.~Giles, C.~B.~Thorn,
{\it A Lattice Approach To String Theory},
Phys.\ Rev.\ D {\bf 16}, 366 (1977).}

\lref\GR{
I.~S.~Gradshteyn, I.~M.~Ryzhik, {\it Tables of Integrals,
Series, and Products}, Academic Press, 2000.}

\lref\WW{
E.~T.~Whittaker, G.~N.~Watson, {\it A Course of Modern
Analysis}, Cambridge University Press, 1927.}

\lref\Ingham{
A.~E.~Ingham, {\it The Distribution of Prime Numbers}, Cambridge
University Press, 1932.}

\Title{\vbox{
\hbox{hep-th/0311231}
\hbox{DAMTP-2003-131}
\hbox{DESY-03-186}
}}
{\vbox{\centerline{On the exact open-closed vertex }
\smallskip
\vbox{\centerline{in plane-wave light-cone string field theory}}}}
\centerline{James Lucietti$^{\, \flat}$, Sakura Sch\"afer-Nameki$^{\, \sharp}$ and Aninda Sinha$^{\, \flat}$ }
\bigskip
\centerline{\it $^\flat$DAMTP, University of Cambridge}  
\centerline{\it Wilberforce Road, Cambridge CB3 OWA, U.K.}
\centerline{\it $^\sharp$II. Institut f\"ur Theoretische Physik,
University of Hamburg} 
\centerline{\it Luruper Chaussee 149, 22761 Hamburg, Germany}
\footnote{}{\tt Email: J.Lucietti, S.Schafer-Nameki, A.Sinha@damtp.cam.ac.uk}

\vskip1.4cm
\centerline{\bf Abstract}
\bigskip
\noindent
The open-closed vertex in the maximally supersymmetric type
  IIB plane-wave light-cone string field
  theory is considered and an explicit solution for the bosonic part of
  the vertex is derived, valid for all values of the mass parameter, $\mu$. This vertex is of relevance to IIB plane-wave
  orientifolds, as well as IIB plane-wave strings in the presence of D-branes and their gauge theory duals. 
Methods of complex analysis are used to develop a systematic procedure
  for obtaining the solution. This
  procedure is first applied to the vertex in flat space and then extended to the plane-wave case. The plane-wave solution for
  the vertex requires introducing certain ``$\mu$-deformed Gamma functions'',
  which are generalizations of the ordinary Gamma function. 
The behaviour
  of the Neumann matrices is graphically illustrated and their
  large-$\mu$ asymptotics are analysed.
\bigskip

\Date{11/2003}


\newsec{Introduction}
\pano

The plane-wave limit of the AdS/CFT correspondence has been a fertile
and active area of research. In the most prominent variant, this limit relates string theory in the
plane-wave background obtained as a Penrose-limit of $AdS_5\times S^5$
\refs{\metsaev, \metsaevtseytlin} with a particular
sector of ${\cal N}=4$, $d=4$ Super Yang-Mills theory \refs{\bmn}. This
sector of the SYM theory is
known as the BMN sector. In this limit both the string theory and the gauge theory are perturbative and thus are independently
accessible to direct computations. 

Understanding this duality in the presence of
interactions is an important problem, as it will provide more underpinning
evidence for the correspondence. In order to study interactions on the
string theory side, one has to resort to light-cone superstring field
theory, as developed initially in \refs{\GSstringfield,\gstwo} for flat space and
extended to the plane-wave string theory in \refs{\spradlinvolovich,\spradlintwo,\klebanovspradlinvolovich,\pankstef}.
When considering plane-waves with D-branes
\refs{\Billo,\Dabholkar, \lee,\nayak, \stone,\bala,\bgg,\sttwo,\gg,\stthree,\oblique} or
orientifolds \refs{\bgmnn,\schnitzer,\asnvs}, open
strings are naturally present and their interactions need to be
taken into account. This open-closed
string theory is captured by seven basic interactions that can occur. 
In addition to the free
closed and open string Hamiltonians, $H_{cc}$ and $H_{oo}$, these
open-closed interactions enter the Hamiltonian in the following manner
\eqn\lssa{
H=H_{cc}+H_{oo}+\sqrt{g_s}(H_{o\leftrightarrow oo}+H_{o\leftrightarrow
  c})+g_s (H_{c\leftrightarrow cc}+H_{o\leftrightarrow
  oc}+H_{oo\leftrightarrow oo}+H_{o\leftrightarrow
  o}+H_{c\leftrightarrow c})\,,
}
where $\sqrt{g_s}$ and $g_s$, denote the open and closed
string coupling constants, respectively.
The first of the $O(\sqrt{g_s})$ terms represents the cubic open string
interaction, while the second represents the open to closed
transition.
The $H_{o\leftrightarrow oo}$ interaction was studied in
\refs{\bogdanone,\chandra} and the large-$\mu$ limit of the open-closed
interaction $H_{o\leftrightarrow c}$ and its gauge theory implications was the focus of \refs{\gomis}.

Each term in the Hamiltonian can be computed in two steps. Firstly,
one imposes the geometrical continuity conditions on the coordinates and
conjugate momenta, \ie\ the kinematical constraints. Secondly, one imposes that the Hamiltonian
satisfies the supersymmetry algebra. The first step of the calculation
involves calculating the so-called Neumann matrices, which follow from
solving the continuity conditions written in terms of the string
modes. These Neumann matrices thus relate the various string modes and form
a crucial ingredient in determining the correction to the
Hamiltonian. The second step involves determining the so-called
prefactor by imposing the supersymmetry algebra.
The prefactor is a polynomial in creation operators, which implements the
dynamical constraints. Determining
the prefactor depends on certain decomposition theorems, which require
the knowledge of the Neumann matrices.

In the plane-wave background
the explicit determination of these Neumann matrices is highly
non-trivial, in particular due to their
dependence 
on the background constant Ramond-Ramond five-form flux $\mu$. As the large-$\mu$ limit of the plane-wave string theory is
conjectured to correspond to the BMN sector of the ${\cal N}=4$
SYM theory,
the Neumann matrices in this limit have been of foremost interest and 
the only example so far of a solution known for all values of $\mu$
has been the cubic
closed string vertex \refs{\spradlinvolovich,\he, \Schwarz}.

In this paper, we will explicitly construct the Neumann matrices for
the open-closed transition vertex, which are valid for all $\mu$. 
In particular, this allows to rigorously obtain both the large-$\mu$
asymptotics as well as to reproduce the correct flat space limit. The corresponding
vertex in flat-space has been discussed in \refs{\GSstringfield, \ShapiroThornone, \ShapiroThorntwo, \Thorn, \GilesThorn,
\GreenWai}.

The open-closed vertex in the large-$\mu$ limit and its relation to
the gauge theory was discussed in \refs{\gomis}. 
Our analysis will determine the solution to the vertex
equations for all values of $\mu$ and then study the asymptotics for large
$\mu$. This stands in contrast to taking the large-$\mu$ approximation of
the vertex equations before solving them, which is what has been
proposed in \refs{\gomis}. 
The expressions obtained in this paper 
will yield, compared to the naive approximation, the same large-$\mu$
asymptotics for one of the matrices, but renormalized results for the
other two. So, one has to treat
such naive approximations with a grain of salt and has to carefully analyse
whether they are mathematically justified, which generically they are
not. This point shall be elaborated upon in due course.

In flat space the open-closed Neumann matrices \refs{\GSstringfield}
are constructed out of
certain functions $u_m$, which are defined as
\eqn\lssc{
u_m={\Gamma(m+1/2)\over \sqrt{\pi} \Gamma(m+1)}\,,
}
where $m$ represents the mode number. It will turn out that in the
plane-wave background these functions are replaced by certain
``$\mu$-deformed'' generalizations; this will in particular require the
definition of two generalizations of the Gamma function, which we
shall refer to as {\it $\mu$-deformed Gamma functions}.

In deriving the Neumann matrices, we will use methods of complex
analysis in order to rewrite certain infinite sums in terms of contour integrals
on the
complex plane, where the complex variable will represent the mode
number that is being summed over. The pole and zero structure of the
Neumann matrices will be motivated using this integral representation
of the sums. We illustrate this method in the flat space case.
However, in the plane-wave case certain subtleties in the method
arise, which are due to the presence of the mode
numbers $\omega_{n}= { \rm sgn} (n) \sqrt{n^2 + \mu^2}$ and the thereby
resulting square root branch cuts. These points will be addressed in detail.

We shall determine the Neumann matrices for both Dirichlet
and Neumann boundary conditions of the open string.
In flat
space, these matrices are related by T-duality. In the plane-wave
background, statements about T-duality are more obscure. In
our case it will turn out that the Neumann matrices for Dirichlet
and Neumann boundary conditions differ by a
$\mu$-dependent factor, which goes to unity in the $\mu=0$ limit.
In this limit, our solutions are precisely equal to the flat space
result. The implications of this need further investigation.

The paper is organized as follows. In section 2, we shall derive the
continuity conditions to be imposed on the open-closed vertex and
thereby derive the equations for the Neumann matrices.
In section 3, we
will illustrate a procedure using methods of contour integration to
solve these constraints explicitly. Using this procedure we will
re-derive the known flat space solutions \refs{\GSstringfield}. We
elaborate on the subtleties arising from branch cuts and branch point singularities in
the plane-wave background and motivate the pole and zero
structure for the Neumann matrices. 
In section 4, we will solve the equations for the Neumann
matrices for all values of $\mu$ in the case of Neumann as well as Dirichlet boundary conditions. This involves the
definition of new $\mu$-deformed Gamma-functions.
In sections 5 and 6, we will analyse the
large-$\mu$ asymptotics of the solutions and their behaviour will be
illustrated graphically.
We conclude with discussions and open problems in section 6.
Appendix A summarizes known identities relevant to the flat space
analysis and an example using the contour
method to sum a series is provided. In appendix B, several key properties and the asymptotics of the newly defined
$\mu$-deformed Gamma functions are discussed.


\newsec{The open-closed vertex}

In this section, we will set up the notation and conventions to be used in
the rest of the paper. We will derive the continuity conditions to be
imposed on the bosonic vertex and the resulting constraint on the
Neumann matrices.

\subsec{Neumann boundary conditions}

We start with the calculation in the plane-wave of the open-closed vertex, with Neumann
boundary conditions on the open string. The corresponding discussion for the flat space superstring can be found in \refs{\GSstringfield}. The bosonic part of the world sheet action is
\eqn\lssd{
\int_0^{\pi |\alpha|}d^2\sigma (\partial X\cdot\partial X+\mu^2 X^2)\,,
}
where we have suppressed the spacetime vector index for
convenience. The length of the world-sheet is parametrized by
$|\alpha|=|2 p^+ \alpha'|$, which without loss of generality, we will
take to be unity for the purpose of this paper\foot{This is also
the convention used in \refs{\gomis}.}. The equations of motion read
\eqn\lsse{
(-\partial_\tau^2+\partial_\sigma^2-\mu^2)X=0\,.
}
The mode expansions for the closed string and open string with Neumann
boundary conditions, which satisfy the above equations of motion are 
\eqnn\lssf
\eqnn\lsslo
$$
\eqalignno{
X^I_{closed}(\sigma, \tau) &= x^I_c\cos \mu\tau + p^I_c{\sin \mu\tau\over \mu}+i 
\sum_{m\not=0} {1\over \omega_{2m}} \left(
\alpha_{m}^I e^{-i(\omega_{2m}\tau+2m\sigma)}
+ \tilde{\alpha}_{m}^I e^{-i(\omega_{2m}\tau-2m\sigma)} \right) \,. &\cr
&& \lsslo \cr
X^I_{open}(\sigma,\tau)& =x_0^I \cos \mu \tau+ p_o^I{\sin \mu \tau \over
\mu}+i\sum_{m\neq 0}{1\over \omega_m}\beta^I_m e^{-i\omega_m \tau}\cos
m\sigma \,, & \lssf
}
$$
which at $\tau=0$ become
\eqn\lssg{\eqalign{
X^I_{closed}(\sigma) &= x^I_c + i 
\sum_{m\not=0} {1\over \omega_{2m}} \left(
\alpha_{m}^I e^{-2im\sigma}
+ \tilde{\alpha}_{m}^I e^{2im\sigma}
 \right) \cr
X^I_{open}(\sigma) &= x^I_o + i \sum_{m=1}^{\infty} {1\over \omega_m}
(\beta_{m}^I- \beta_{-m}^I) \cos {m \sigma} \,.
}}
We also define $P^I={\partial\over\partial\tau}X^I$ so that at $\tau=0$,
\eqn\lssh{\eqalign{
P^I_{closed}(\sigma) &= p^I_c+ 
\sum_{m\not=0} \left(
\alpha_{m}^I e^{-2im\sigma}
+ \tilde{\alpha}_{m}^I e^{2im\sigma}
 \right)\cr
P^I_{open}(\sigma) &= p^I_o +  \sum_{m=1}^{\infty} 
(\beta_{m}^I+ \beta_{-m}^I) \cos {m \sigma}  \,.
}}
The non-trivial commutation relations are
\eqn\lssi{\eqalign{
\left[\beta_m,\beta_n\right]&=\omega_m \delta_{m,-n}\, \cr
\left[\alpha_m,\alpha_n\right]&={\omega_{2m}\over 2}\delta_{m,-n}\, \cr
\left[{\tilde \alpha_m},{\tilde \alpha_n}\right]&={\omega_{2m}\over 2}\delta_{m,-n}\,.
}}
The conventions have been chosen such that
\eqn\omdef{
\omega_n ={\rm sgn}(n) \sqrt{n^2 + \mu^2} \,,
}
where $\mu$ is the RR-field strength and we have parametrized the string world-sheet to be of length $\pi$. Note that the mode expansion and the commutation relations have the expected flat space limit and coincide with the expressions given in \refs{\GSstringfield}.
For the open-closed vertex, the following relation has to be satisfied
by the fields at $\tau=0$
\eqn\lssj{
X^I(\sigma)_{open} = X^I(\sigma)_{closed} \,, \qquad 
P^I(\sigma)_{open} + P^I(\sigma)_{closed}=0 \,, 
}
which results in terms of the modes in 
\eqnn\eqone
\eqnn\eqtwo
$$
\eqalignno{
{1\over \omega_{2n}} (\alpha_n - \alphatilde_{-n})
- \sum_{m=1}^{\infty} {1\over \omega_m} c_{nm}(\beta_m - \beta_{-m}) &= 0 
&\eqone\cr
(\alpha_n + \alphatilde_{-n}) + \sum_{m=1}^{\infty} c_{nm} (\beta_m +
\beta_{-m}) &= 0  \,. &\eqtwo
}
$$
These equations are understood as holding upon the vertex $|V\rangle$
and the constants $c_{nm}$ are given by
\eqn\lssk{
c_{nm} ={1\over \pi} \int_0^{\pi}
d\sigma \, e^{ 2in\sigma} \cos(m\sigma) \,=\,
\left\{
\eqalign{
\half (\delta_{m,-2n} + \delta_{m,2n}) & \qquad\qquad m \hbox{ even,} \cr
{4in \over \pi (4n^2-m^2)} & \qquad \qquad m \hbox{ odd}\,.
}\right.
}
The relations \eqone\ and \eqtwo\ are equivalent to the
following conditions (derived by combining the $n$ and $-n$ versions of the equations)
\eqn\glueone{
\alpha_n + \alphatilde_n +\beta_{-2n} =0\,,  
}
and
\eqn\gluetwo{
\alpha_n - \alphatilde_n - {i n\over \pi}\sum_{m=0}^{\infty}
{1\over (m+1/2)^2 - n^2} \left( 
\left(1- {\omega_{2n}\over \omega_{2m+1}}\right) \beta_{2m+1} + 
\left(1+ {\omega_{2n}\over \omega_{2m+1}}\right) \beta_{-2m-1}
\right) = 0 \,. 
}
These equations reduce precisely to the relations in
\refs{\GSstringfield} for $\mu=0$\foot{Note that there is a
slight convention mismatch in \refs{\GSstringfield}, in that the mode
expansion in (2.11) in \refs{\GSstringfield} together with the
definition of the $c_{nm}$ do not give rise to (7.10,11). This can
be remedied by choosing the opposite sign for $\sigma$ in the
exponentials in (2.11).}.
\glueone\ implies that the vertex $|V\rangle= \exp(\Delta)
|\Omega\rangle$ has to have a decomposition $\Delta =\Delta_1 +
\Delta_2$, where
\eqn\vertexone{
\Delta_1 = -\sum_{m=1}^\infty {\sqrt{2}\over \omega_{2m}} \beta_{-2m}
\alpha^I_{-m}\,,
}
and where we defined $\sqrt{2}\alpha^{I/II}= \alpha \pm
\alphatilde$, so that $\alpha_0^{II}=0$.
In order to solve \gluetwo, we further make the ansatz
\eqn\deltatwo{
\Delta_2 = \sum_{m,n=0}^\infty A_{mn} \beta_{-2m-1} \alpha^{II}_{-n} + \half B_{mn} \beta_{-2m-1}\beta_{-2n-1} + \half C_{mn} \alpha_{-m}^{II} \alpha_{-n}^{II} \,.
}
The resulting generalizations of the equations (7.15)-(7.18) in
\refs{\GSstringfield}\foot{Note that there is a factor of 2 missing on the LHS of
    equation (7.15) in \refs{\GSstringfield}.} are
\eqnn\tosolveone
\eqnn\tosolvetwo
\eqnn\tosolvethree
\eqnn\tosolvefour
$$
\eqalignno{
-{2\sqrt{2} i n\over \pi} \sum_{m=0}^\infty
{1\over 
\omega_{2m+1}- \omega_{2n}}
A_{mk} =&\, \delta_{n,k}  &\tosolveone\cr
-\sum_{m=0}^\infty {B_{mk}\over \omega_{2m+1} - \omega_{2n}}
=& {1\over (\omega_{2n} +\omega_{2k+1})\omega_{2k+1}} &\tosolvetwo \cr
{4\sqrt{2} i n\over \pi \omega_{2n}} \sum_{m=0}^{\infty} { 1\over \omega_{2m+1}+ \omega_{2n}} A_{mp}=&\, C_{np} &\tosolvethree \cr
{4\sqrt{2} i n\over \pi\omega_{2n}} \left(\sum_{m=0}^\infty {1\over \omega_{2m+1}+ \omega_{2n}}
B_{mp} + {1\over (\omega_{2p+1}-\omega_{2n}) \omega_{2p+1}}\right)=&\,
A_{pn}\,.
&\tosolvefour
}
$$
The fact that the open string must join smoothly at its
end points imposes the additional condition for $A$ that
\eqn\tosolvefive{
\sum_{m=0}^\infty A_{mk}=0\,,
}
and another condition for $B$
\eqn\tosolvesix{
\sum_{m=0}^\infty B_{mk}={1\over \omega_{2k+1}}\,.
}
For $\mu\rightarrow 0$ these reduce to the correct flat space relations.
When we consider the limit $\mu\rightarrow \infty$, naively from
equation \tosolveone\ it seems that $A$ must be of order $O(1/\mu)$ and from 
equations \tosolvetwo, \tosolvethree\ that $B,C$ must be
$O(1/\mu^3)$. Considering equation \tosolvefour\ and
neglecting the $B$ term, we see that the large-$\mu$ asymptotics of
$A$ by this naive analysis is given by
\eqn\Aasymp{
A_{pn} ={1\over \pi \mu}{ 8 i n \sqrt{2} \over(2p+1)^2-(2n)^2} + O\left({1\over
\mu^3}\right)\,.
}
As we will see in the section on the large-$\mu$ asymptotics, the
naive approximations differ for $B$ and $C$ from the actual results.

\subsec{Dirichlet boundary conditions}

In the case of Dirichlet boundary conditions, the open string mode
expansion is
\eqn\Dirichletmode{
X^I_{open}(\sigma,\tau)= ({\rm zero\;modes}) + \sum_{m\neq 0}{1\over \omega_m}\beta^I_m
e^{-i\omega_m \tau}
\sin m\sigma\,.
}
The presence of zero-modes is dependent on what type of D-brane is
being considered. In particular, for class I branes, the zero modes
vanish, however for class II and oblique branes, the zero modes are in fact
$\sigma$-dependent \refs{\bgg, \gg, \oblique, \sttwo}\foot{In particular, for the class I $D7$-brane, considered in \refs{\gomis}, there
are no zero-modes for the open string.}.
For class I branes the vertex equations take the form
\eqnn\Deqone
\eqnn\Deqtwo
$$
\eqalignno{
i {1\over \omega_{2n}} (\alpha_n - \alphatilde_{-n})
- \sum_{m=1}^{\infty} {1\over \omega_m} \check{c}_{nm}(\beta_m +\beta_{-m}) &= 0 
&\Deqone\cr
i (\alpha_n + \alphatilde_{-n}) + \sum_{m=1}^{\infty} \check{c}_{nm} (\beta_m -
\beta_{-m}) &= 0  \,, &\Deqtwo
}
$$
where
\eqn\ccheck{
\check{c}_{nm}
= {1\over \pi} \int_0^{\pi}
d\sigma \, e^{2in\sigma} \sin(m\sigma) \,=\,
\left\{  
\eqalign{
{i\over 2} (\delta_{m,2n} - \delta_{m,-2n}) &\qquad\qquad m \hbox{ even,} \cr
-{1\over \pi} {2m \over  (4n^2-m^2)} & \qquad\qquad m \hbox{ odd}\,,
} \right.
}
which correspond to the Fourier modes of $\sin (m \sigma)$ that
appear in the Dirichlet open string. 
The resulting vertex equations are then
\eqnn\Dglueone
\eqnn\Dgluetwo
$$
\eqalignno{
\alpha_n - \tilde\alpha_n -\beta_{-2n} &=0 &\Dglueone \cr
\alpha_n + \tilde\alpha_n -i  {2\over \pi} 
\sum_{m=0}^{\infty}
{(2m+1)\over\omega_{2m+1}}\left( 
{ \beta_{2m+1} \over (\omega_{2n}+\omega_{2m+1})} 
+
{\beta_{-2m-1} \over (\omega_{2n}- \omega_{2m+1})}
\right) &= 0 \,.& \Dgluetwo
}
$$
In general, the zero modes for the Dirichlet open strings, for instance
the D-instanton of \refs{\gg, \sttwo}, are $\sigma$-dependent. This leads to a subtlety in the continuity condition since
the Fourier modes of these terms will be non-vanishing for the
non-zero modes. One can take this into account by redefining
$\alpha^I$ and $\alpha^{II}$ in the following manner
\eqnn\newalphaone
\eqnn\newalphatwo
$$
\eqalignno{
\left(\alpha^I_{n}\right)_{new}&=\left(\alpha^I_{n}\right)_{old}+{i\over
2\sqrt{2}}\omega_{2n}(f_n+f_{-n})&\newalphaone \cr
\left(\alpha^{II}_{n}\right)_{new}&=\left(\alpha^{II}_{n}\right)_{old}+{i\over
2\sqrt{2}}\omega_{2n}(f_n-f_{-n})\,,&\newalphatwo}
$$
where
\eqn\fdef{
f_n={1\over \pi}\int_0^{\pi}d\sigma \, {(\rm zero~modes)}\, e^{2in\sigma}\,.
}
With this redefinition, one would need to define the vacuum in terms
of these new oscillators\foot{Explicitly, if $(\alpha_n)_{new} =
(\alpha_n)_{old} + c_n$, where $c_n$ are $c$-numbers, then $|\Omega
\rangle_{new} = \exp \left( -\sum_1^{\infty} 2 { c_n \over \omega_{2n}} (\alpha_{-n})_{old}
\right) | \Omega \rangle_{old}$. Clearly then $(\alpha_n)_{new} |\Omega
\rangle_{new} =0$ for $n>0$. }. The analysis will now be the same as that
for the case of vanishing zero modes.
 
The key point is again to solve for the part of the vertex involving odd open string
modes. 
The ansatz for the vertex is
\eqn\Ddeltatwo{
\check{\Delta}_2 = \sum_{m=0,n=1}^\infty \check{A}_{mn} \beta_{-2m-1}
\alpha^{I}_{-n} + \half \sum_{m,n=0}^\infty
\check{B}_{mn} \beta_{-2m-1}\beta_{-2n-1} + \half \sum_{m,n=1}^\infty \check{C}_{mn} \alpha_{-m}^{I} \alpha_{-n}^{I} \,,
}
for which \Dgluetwo\ imposes the following conditions
\eqnn\Dtosolveone
\eqnn\Dtosolvetwo
\eqnn\Dtosolvethree
\eqnn\Dtosolvefour
$$
\eqalignno{
i {\sqrt{2} \over \pi} \sum_{m=0}^\infty
{(2m+1)\over \omega_{2m+1}- \omega_{2n}} \check{A}_{mk}
=&\, \delta_{n,k} & \Dtosolveone \cr
\sum_{m=0}^\infty {(2m+1)\over \omega_{2m+1} - \omega_{2n}}\check{B}_{mk}
=& {(2k+1)\over (\omega_{2n} +\omega_{2k+1})\, \omega_{2k+1}}
&\Dtosolvetwo \cr 
i {2 \sqrt{2} \over \pi \omega_{2n}} \sum_{m=0}^\infty {(2m+1)\over
\omega_{2m+1}+ \omega_{2n}} \check{A}_{mk}=& \, \check{C}_{nk} &\Dtosolvethree\cr
i { 2\sqrt{2} \over \pi \omega_{2n}} \left(\sum_{m=0}^\infty {(2m+1)\over \omega_{2m+1}+ \omega_{2n}}
\check{B}_{mk} - {(2k+1)\over (\omega_{2k+1}-\omega_{2n})
  \omega_{2k+1}}\right) =&\,  \check{A}_{kn}\,. & \Dtosolvefour
}
$$
$(X(0)-X(\pi))|V\rangle=0$ is autmatic for the non-zero modes and so
does not impose additional constraints.
For $\mu\to 0$ these reproduce the flat space equations of
\refs{\GreenWai}. Again, one can study a naive approximation of the
solutions as $\mu\rightarrow\infty$, which yields, \eg,
\eqn\Acheckasymp{
\check{A}_{kn} = -{4i\sqrt{2} (2k+1)\over \pi ((2k+1)^2 -(2n)^2)\mu} +
O\left({1\over \mu^3}\right)\,.
}
We shall determine the solutions and in particular
this will show that they are closely related by a $\mu$-dependent
factor to the solutions in the Neumann case.


\newsec{Summation technique for solving the vertex equations}

\subsec{Contour integration method}

Let us consider the system of equations \tosolveone -\tosolvesix\ in flat
space, \ie\ when $\mu=0$. In order to try and solve these we employ
the following technique. Let $f(z)$ be analytic except for possibly
poles, which will be at positions $z_k$. Suppose $f(z)$ is zero, when
$z$ is a negative integer. Then by Cauchy's theorem we
have
\eqn\contone{
\sum_{n=0}^{\infty} f(n)+ \sum_k Res_{z=z_k} \pi\cot(\pi z)f(z) =
\lim_{R\to \infty} \oint_{C_{R}} {dz\over 2\pi i} \pi\cot(\pi z) f(z)\,,
}
where $C_R$ is the contour given by a circle of radius $R$ centred
on the origin, which does not intersect any poles of the integrand (so
in particular $R \neq 1,2,3...$). The contour is depicted in figure 1.

\fig{Contour $C_R$.}{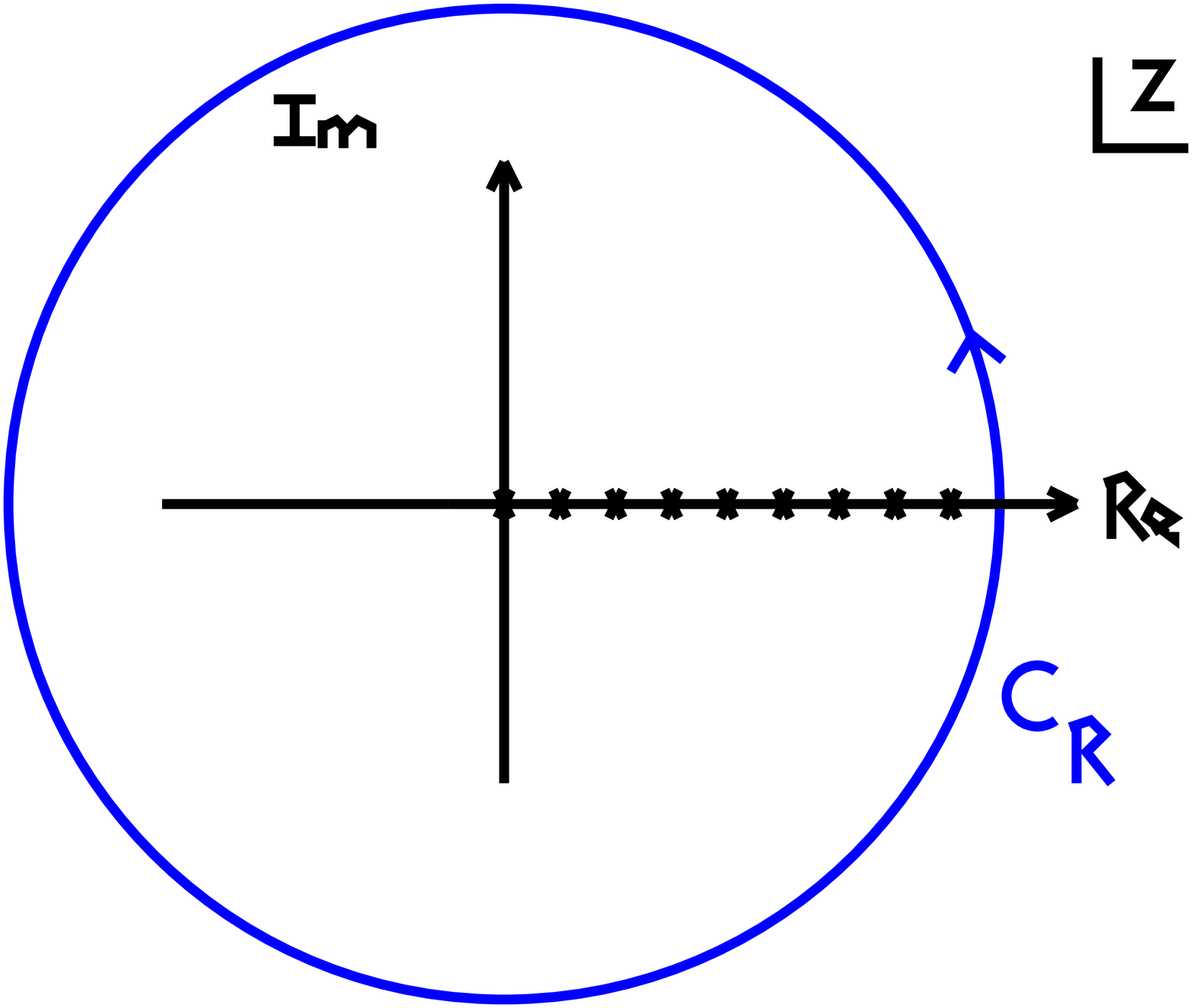}{2.2truein}

In order for our technique to
be useful we require that the RHS of the above equation vanishes,
which leaves us with an expression for $\sum_0^{\infty} f(n)$. This
will impose a condition on the behaviour of $f(z)$ at infinity, which
we will now deduce. Fortunately this turns out to be easy since
$|\cot(\pi z)|$ is bounded on $C_R$ as $R \to \infty$. Therefore it is
sufficient to require that $|zf(z)| \to 0$ on $C_R$ as $R \to
\infty$. Thus the problem can be reformulated as: given $\sum_0^{\infty} f(n)$
find an $f(z)$ which has the correct asymptotic behaviour. This will
then provide us with a solution for $f(n)$. At this stage it is not
obvious under what conditions the solutions are unique and we shall
return to this point below. It is important to note that if $f(z)$ were
to have no
poles, it would be analytic everywhere and thus due to the asymptotic
property would be bounded; hence by Liouville's theorem a constant,
which must equal zero. Thus we conclude that $f(z)$ must have poles. An explicit
example of this method is provided in appendix A.2 for proving a well
known identity.

Note that given the behaviour at infinity, the pole and
zero structure (\ie\ positions and multiplicities of all the poles and
zeroes) may be sufficient to determine $f(z)$ uniquely. To do this we
employ deeper results on holomorphic functions, namely, if $h(z)$ is
analytic everywhere in $\Cop$, has no zeroes and is of
{\it order} zero, then it is a constant. An analytic function is of
order zero, if $\log M(R) = O(R^{\epsilon})$ for $\epsilon >0$, where
$M(R)$ is the maximum of $|h(z)|$ on circles of radius $R$ (centred on
the origin). We may now apply this to our case. Suppose we have two
possibilities for $f(z)$, call them $f_1(z)$ and $f_2(z)$, with the
same pole and zero structure. Consider $f_1(z)/f_2(z)$. This is
analytic everywhere with no zeroes. If $f_2(z) \sim z^{-\beta}$ where
$\beta >1$ then $f_1(z)/f_2(z)=O(z^k)$ where $k \geq \beta -1$. Thus
we see that for $f_1/f_2$ we have $\log M(R) = O(\log R)$ and hence it
is of order zero. We deduce that $f_1/f_2 = A$, a constant. Techniques
of complex analysis similar to these used above can be found, \eg, in
\refs{\Ingham}.

\subsec{Flat space case}

As a warm-up for the plane-wave case, we shall use the above 
contour method to explicitly construct the known solutions to the flat space
open-closed vertex equations. These were obtained in 
\refs{\GSstringfield} by using the identities given in appendix A. 
Firstly, consider the flat space limit $\mu=0$ of the equations
\tosolveone -\tosolvesix\ for $A$ and $C$ 
\eqnn\Aone
\eqnn\Atwo
\eqnn\Athree
$$
\eqalignno{ 
\sum_{m=0}^{\infty} A_{mn}&=0, &\Aone\cr
\sum_{m=0}^{\infty} {A_{mk}\over 2m+1-2n} &=
-{\pi\delta_{nk}\over 2\sqrt{2} in}\,, &\Atwo\cr
{i\sqrt{2} \over \pi}\sum_{m=0}^{\infty} { A_{mp}\over{m+1/2+n}}&=
C_{np}\,. &\Athree
}
$$
In order to solve these equations, we first convert the sums into 
contour integrals using the method of the previous section, \ie, by introducing the complex function
$A(m,k)$ of $m \in \Cop$ such that it
coincides with $A_{mn}$, when $m$ is a non-negative integer. The idea will be to determine the
poles and zeroes of $A$ and reconstruct the function using this
information. For example the contour integral for the LHS of \Atwo\ takes the form
\eqn\contintA{
\oint_{C}dm\,\cos(\pi m) \Gamma(-m)\Gamma(m+1){A(m,k)\over{m+1/2-n}}\,,
}
where $C$ is the contour which encloses only the positive integers. We
will send the contour to infinity assuming that $A$ has the asymptotic
behaviour, such that the integrand tends to zero at infinity. 
From \Aone\ we can deduce that $A(m,k)$ will have at most simple
poles at half integer values of $m$ (since they will cancel with the
zeroes of $\cos(\pi m)$), if we assume no relative cancellations
between residues of $A(m,k)$ \foot{In fact it seems that this
must be the case, for  $A_{mk}$ must satisfy two separate sums and
thus if the residues cancelled in one they would not in the other.}. Also we see that $A(m,k)$ must have
zeroes at negative integers (again assuming no relative
cancellations occur) since otherwise $\Gamma(m+1)$ would give non-zero
contributions. Further, setting $n=k$ in \Atwo\ implies that there
must be a simple pole at $m=k-1/2$. 
Thus $A(m,k)$ must be of the form
\eqn\Aflat{
A(m,k)={f(m,k)\over{m+1/2-k}}\,,
}
where $f(m,k)$ is inversely proportional to $\Gamma(m+1)$.
Now consider equation \Athree. If $f(m,k)$ had no further poles,
then following the line of reasoning above, $C_{np}$ would vanish,
which is ruled out by physical considerations. Thus $f(m,k)$ must have
further poles at $m=-n-1/2$, for each $n=0,1,\cdots$. Equation \Aone\ immediately tells us that these should be
simple poles. Thus $f(m,k)$ must be proportional to
$\prod_{n=0}^{\infty}{1/(m+n+1/2)}$. The product is divergent as it
stands, but can be made convergent by using the Weierstrass
form of the Gamma function as in appendix A. Thus $f(m,k)$ is proportional to
$\Gamma(m+1/2)/\Gamma(m+1)$, which in the notation of
\refs{\GSstringfield} is defined as $\sqrt{\pi}u_m$. As there cannot be
further zeroes or poles in $A(m,k)$, we have determined $A(m,k)$ up to
an unknown function in $m$, with neither zeroes nor poles. Using the theorem
in the previous section we can show that if this function is
holomorphic and has the required asymptotic behaviour for the contour
method to work, then it must be a constant. The constant of
proportionality can now be easily determined by considering equation
\Atwo\ with $n=k$. The complete answer for $A_{mk}$ is thus
\eqn\Aflatagain{
A_{mk}=i\sqrt{2}{u_m u_k\over 2m+1-2k}\,.
} 
We should emphasise that we have proven that the above expression for
$A_{mk}$ is the only meromorphic function in $m$ with the given poles and
zeroes, which solves \Aone\ and \Atwo\ with the forementioned
asymptotic property.
Using this solution in equation \Athree, we can now easily derive
the solution for $C_{nk}$ using the contour integration technique. The
non-trivial residues of relevance in the integration are at
$m=-n-1/2$. The solution works out to be
\eqn\Cflat{
C_{nk} = {u_n u_k\over n+k} \,.
}
In a manner that is identical to deriving the solution for $A_{mk}$ as
illustrated above, it can be shown that the solution for $B_{mk}$ is
given by
\eqn\Bflat{
B_{mk}={u_m u_k \over 2m+2n+2}\,.
}
The so-derived solutions for $A,B,C$ are precisely as given in the
literature \refs{\GSstringfield}.

\subsec{Subtleties in the plane-wave case}

We now turn to the plane-wave open-closed vertex equations and their
solutions. This section provides a discussion of various subtleties, which arise in generalizing the contour method to the
plane-wave case. The reader interested only in the solution to the
vertex equations may therefore turn directly to the next
section, where the full solution will be presented and proven.

The main strategy in solving \tosolveone -\tosolvesix\ will
be to proceed as in the flat space case, \ie, to analyze the structure
of poles and zeroes. Again, we assume that there are no relative
cancellations of residues. 
Consider for definiteness the equation for $B_{mk}$, \tosolvetwo. The relevant
contour integral is
\eqn\subtleB{
\oint dm\, \cot(\pi m) {B(m,k)\over \omega_{2m+1} -\omega_{2n}} \,.
}
In order to satisfy \tosolvetwo\ for all values of $n$, $B(m,k)$ 
cannot have poles in $m$ at positive half-integers. This can be seen
as follows. \tosolvesix\ implies that $B$ has to have a pole at
one negative integer value. The integral for \tosolvetwo\
obtains precisely a residue at this integer as well (since the only extra
factor in the integrand compared to \tosolvesix\ has a pole at $m=n-1/2$) and the value of the residue is precisely
the LHS of \tosolvetwo. Now, assume $B$ had a pole for a positive half-integer, $p+1/2$. If the
denominator term $1/(\omega_{2m+1} -\omega_{2n})$ has no other poles
at $p+1/2$, then due to the $\cot$-factor the corresponding
residue vanishes. However, since \tosolvetwo\  has to hold for all
$n$, for $p=n$, there would be an additional non-zero residue. So
we conclude that $B$ cannot have poles at positive half-integer values.
However, $B$ can have poles
at negative half-integers, as these are again cancelled by the
$\cot(\pi m)$ factor; in fact it must due to \tosolvefour\ .
 Further $B$ has to
have zeroes in $m$ at negative integers (cancelling the poles of the $\cot
(\pi m)$), except for $m=-k-1$, where there has
to be a non-trivial residue of the above integral, which gives rise to
the RHS of \tosolvetwo. This yields the following ansatz 
\eqn\guessB{
B (m,k) = {g(m,k)\over (\omega_{2m+1} +\omega_{2k+1})} \,,
}
where $g(m,k)$ has poles at negative half-integers and zeroes at
negative integers. From these constraints alone one may be led to
choose $g(m,k)= u_m u_k$, however at this point the following
subtlety, characteristic for the plane-wave case, presents itself.

The key problem arises through contributions to the contour integral
from the square root branch cut of $\omega_{2m+1}$.
The branch points are located at $m_{\pm}= -\half \pm i {\mu\over 2}$.
Writing $\omega_{2m+1}= \sqrt{2m+1 - i \mu}\sqrt{2m+1 + i \mu}$, 
and choosing the cuts for the square root factors to extend from $m_-$
to $i \infty$ and $m_+$ to $i \infty$, respectively, we obtain a branch line extending from $m_-$
to $m_+$. This choice of cut is suitable, as it ensures that when
restricted to the integers, $sgn(n)$ is automatically incorporated in
$\omega_n$, \ie, to the right of the cut the phase of the square root
is chosen to be $+1$ and to the left (in particular for all
negative integers) it is $-1$.
The important point to note now, is that in the contour method, one has
to take the contributions from the integral around the cut (depicted
blue in figure 2) into account. These come from
the two line integrals along $L_1$ and $L_2$, as well as the integrals
around the branch points, $K_{1,2}$. We have also set $\mu_{\pm}= \pm
i \mu/2$. 

\fig{Contribution from the branch cut.}{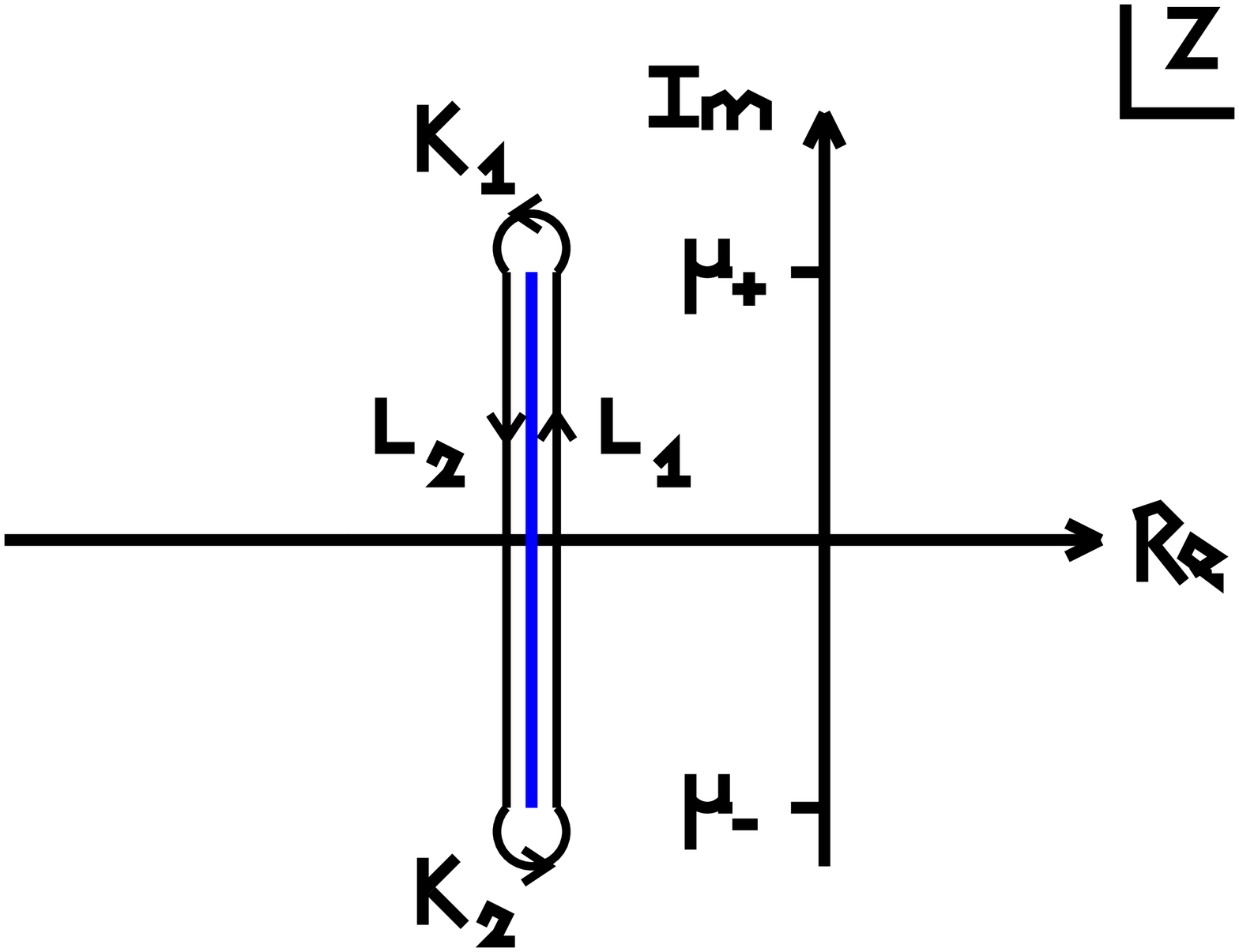}{2.2truein}

In crossing the branch cut, the square root picks up
the following phases. Along $L_1$ the argument of the $\sqrt{2m+1+i\mu}$-factor is $\pi/4$ and of
the $\sqrt{2m+1-i\mu}$ is $-\pi/4$, whereas along $L_2$ the arguments are $\pi/4$ and
$3\pi/4$. 
 Thus,
by crossing from $L_1$ to $L_2$, $\omega_{2m+1}$ picks up a minus
sign. In particular, a term
$$
\int_{L_1} dm [...] {1\over (\omega_{2m+1}+\omega_{2n})(\omega_{2m+1}+\omega_{2k})}
\,,
$$
will go over into
$$
\int_{L_2} dm [...] {1\over (-\omega_{2m+1}+\omega_{2n})(-\omega_{2m+1}+\omega_{2k})}
\,.
$$
The most natural way to circumvent this problem is to demand that each
of the line integrals above vanish individually. In order for this to
happen, note that $\cot \pi \left(-{1\over 2} + i{y\over 2}\right)$ is
an odd function in $y\in \Rop$, thus if in the remaining part of the integrand, the
dependence on $m$ led to an even function in Im$(m)$ then the integral
would vanish along each branch. This is possible if the $\mu$ and $m$
dependence was packaged together in the form $\omega_{2m+1}$. This is
natural in light of the plane-wave mode expansions. One can also
anticipate that whenever an index couples to a closed string mode
$\alpha_k$, the $k$ and $\mu$ dependence is packaged into
$\omega_{2k}$. With $m=-1/2+iy/2$, we see that $\omega_{2m+1}$ becomes
$\pm \sqrt{-y^2+\mu^2}$, which is even in $y$. 
In addition one has to ensure that the
contour integrals around the two branch points, $m_{\pm}$, denoted by
$K_{1,2}$ in figure 2, vanish as well. The solutions that we shall
present in the next section will be shown to
satisfy both these requirements. 


\newsec{Solution for the plane-wave open-closed vertex}

\subsec{Neumann boundary conditions}

In order to solve the plane-wave vertex equations, the following
generalized Gamma-functions will be of key importance. We define
\foot{
In version 1 and 2 of the preprint, the ``zero-mode'' part of $\Gamma^I$ in the case of
Neumann boundary conditions was erroneous, as it would have resulted in the non-vanishing
contributions from the branch cuts.}   
the
{\it $\mu$-deformed Gamma-functions of the first and second kind} \foot{To the best of our
    knowledge, these Gamma functions are not related to the
    $q$-deformed Gamma functions and have not been previously
    investigated in the literature.}
\eqnn\muGammaI
\eqnn\muGammaII
$$
\eqalignno{
\Gamma^{I}_{\mu} (z)
&= e^{-\gamma \omega_{2z}/2}\left( {1\over z}\right) 
\prod_{n=1}^\infty \left({\omega_{2n}\over \omega_{2z}+\omega_{2n}} \,
e^{\omega_{2z}/ 2n}\right) 
&\muGammaI \cr
\Gamma^{II}_{\mu} (z)
&=  e^{-\gamma (\omega_{2z-1}+1)/2} \left({2\over \omega_{2z-1}+ \omega_{1}}\right)
\prod_{n=1}^\infty \left({\omega_{2n}\over
\omega_{2z-1}+\omega_{2n+1}}\, e^{(\omega_{2z-1}+1)/ 2n}
\right) \,,&\muGammaII
}
$$
which we shall abbreviate by $\Gamma^I, \Gamma^{II}$, if this does not
cause any ambiguities. By comparison with the Weierstrass form of the
standard Gamma function (see appendix B), these satisfy 
\eqn\gammaflat{
\Gamma^I_{\mu=0}(z)= \Gamma(z)\,,\qquad
\Gamma^{II}_{\mu=0}(z)=\Gamma (z) \,.
}
Various properties of these $\mu$-deformed Gamma-functions are
discussed in appendix B.
These modified Gamma-functions satisfy generalized reflection
identities 
\eqnn\reflecone
\eqnn\reflectwo
$$
\eqalignno{
\Gamma^I_{\mu}(z) \Gamma^I_{\mu}(-z) &= -{\alpha\over z \sin(\pi z)} &\reflecone \cr
\Gamma^{II}_{\mu}(1+z) \Gamma^{II}_{\mu}(-z)&= -{\alpha\over  \sin(\pi
z)}\,, &\reflectwo
}
$$
where $\alpha= 2 \sinh( \pi \mu /2) / \mu$ and $\alpha \to \pi$ as
$\mu \to 0$. 

Further we define the generalizations of the functions 
\eqn\ums{
u_{m} = {\Gamma (m+1/2) \over \sqrt{\pi}\, \Gamma (m+1)} \,,
}
which appear in the flat space solutions of the vertex. Let
\eqnn\vone
\eqnn\vtwo
$$
\eqalignno{
v_m^I &= {(2m+1) \over \omega_{2m+1}} \  
{\Gamma^I (m+1/2) \over \sqrt{\pi}\; \Gamma^{II}(m+1)} &\vone \cr
v_m^{II} &= {2\over \omega_{2m}}\  { \Gamma^{II}(m+1/2) \over \sqrt{\pi}\;
\Gamma^I (m)} \,,&\vtwo 
}
$$
which both reduce to $u_m$ in the flat space limit $\mu\rightarrow
0$. Note that $v^I_z$ has branch points at $-1/2\pm i\mu /2$, whereas
$v^{II}_z$ has branch points at $ \pm i\mu /2$.
Invoking the reflection formulae for the modified Gamma functions, we
compute 
\eqn\voneres{
Res_{m=-n-1/2} \; v_m^I = v^{II}_n /\pi \,.
}
Note that the $\mu$-dependent
constant $\alpha$ cancels.

First, we summarize the solutions to the vertex equations
with Neumann boundary conditions, \tosolveone -\tosolvesix, and then provide the proofs
thereof. The coefficients for the open-closed vertex are given by
\eqnn\Avertex
\eqnn\Bvertex
\eqnn\Cvertex
$$
\eqalignno{
A_{mk} &= i\sqrt{2}  {v^I_m v^{II}_k \over (\omega_{2m+1} -
\omega_{2k})}&\Avertex \cr
B_{mk} &= {v^I_m v^I_k \over (\omega_{2m+1} +
\omega_{2k+1})}&\Bvertex \cr 
C_{mk} &=2 { v^{II}_m v^{II}_k \over (\omega_{2m} + \omega_{2k}) }
\,.&\Cvertex 
}
$$
These solutions can be motivated by noting that the new functions $v^{I}$
and $v^{II}$ have the same pole and zero structure as the $u$ functions in
flat space, to which they further reduce in the $\mu=0$ limit. Moreover, according to
the observation in the previous section, there are no contributions from
the line integrals around the branch cut. This relies on the fact that
$v^I_{-1/2+iy}$ is an even function in $y$. We will also demonstrate that
the branch point singularities in $v^{I}$ and $v^{II}$ will not affect the
calculation.

In order to use the contour method discussed in section 3.1, it is
also crucial to show that $A,B,C$ have the correct asymptotics. Using
the results derived in appendix B.3, we see that $A,B,C$ are all
$O(1/z^{3/2})$ which is the same as in flat space. Thus we are
indeed justified to use the
contour method.
Note also, that by comparison with \refs{\GSstringfield} these
solutions have the correct flat space behaviour.

To prove the assertions, we first consider $B_{mk}$. 
Using the summation technique described previously, we easily see that,
\eqn\Bres{
\sum_{m=0}^{\infty} {B_{mk}\over \omega_{2m+1} - \omega_{2n}} = -
Res_{m=-k-1} {\pi \cot(\pi m) B_{mk}\over \omega_{2m+1} - \omega_{2n}}.
}
A direct computation yields
\eqn\calcone{\eqalign{
&Res_{m=-k-1} {\pi \cot(\pi m) B_{mk}\over \omega_{2m+1} -
\omega_{2n}}  \cr
&= -{\pi v_k^I \over \omega_{2k+1}+\omega_{2n}} \;
\lim_{m \to -k-1} \left(\cot(\pi m)v_m^I \right) \; Res_{m=-k-1}\left(
{1\over \omega_{2k+1}+\omega_{2m+1}}\right) \cr
&=  {1\over (\omega_{2k+1}+\omega_{2n}) \omega_{2k+1}},
}}
where the last equality follows after using the reflection identities
for both types of Gamma function, one of which is required to evaluate
the limit $\lim_{m \to -k-1} [\cot(\pi m)v_m^I ]$. 

Using the solution for $B_{mk}$ as well as \tosolvefour, the
solution for $A_{mk}$ can now be determined. First, note that
\eqn\calctwo{
\sum_{m=0}^{\infty} {B_{mk}\over \omega_{2m+1} +\omega_{2n}} = -
Res_{m=-k-1} {\pi \cot(\pi m) B_{mk}\over \omega_{2m+1} +
\omega_{2n}} - Res_{m=-n-1/2} {\pi \cot(\pi m) B_{mk}\over \omega_{2m+1} +
\omega_{2n}}\,.
}
We see that the first term on the RHS is essentially identical to the
previous calculation and thus we have
\eqn\calcthree{
Res_{m=-k-1} {\pi \cot(\pi m) B_{mk}\over \omega_{2m+1} +
\omega_{2n}} = {1\over (\omega_{2k+1}-\omega_{2n}) \omega_{2k+1}}\,.
}
The second term can be computed as follows
\eqn\calcfour{\eqalign{
 Res_{m=-n-1/2} {\pi \cot(\pi m) B_{mk}\over \omega_{2m+1} +
\omega_{2n}}
& = {v_k^I \pi\over \omega_{2k+1} - \omega_{2n}} \;
Res_{m=-n-1/2} {\cot(\pi m) v_m^I\over \omega_{2m+1} + \omega_{2n}}
\cr
&= \left({v_k^I\,  \pi\over \omega_{2k+1} - \omega_{2n}}\right) \left( {-\pi
\omega_{2n}\over {4n}}\right) \; Res_{m=-n-1/2} v_m^I   \cr
 &=  \left({v_k^I\, v_n^{II}\over \omega_{2k+1} - \omega_{2n}}\right) \left( {-\pi
\omega_{2n}\over {4n}}\right) \,.
}}
Thus, defining
\eqn\calcfive{
A_{mk} = \sqrt{2} i \,  \left({v_m^I\, v_k^{II}\over \omega_{2m+1} - \omega_{2k}}\right)\,,
}
the equation \tosolvefour\ is satisfied. Now we check the
remaining equations satisfied by $A_{mk}$. Since,
$\cot(\pi m)A_{mk}$ only has poles at $m=0,1,2,\cdots$, we see immediately that
\eqn\calcsix{
\sum_{m=0}^{\infty} A_{mk} = 0.
}
To check the other equation we note that,
\eqn\calseven{
\sum_{m=0}^{\infty} {A_{mk}\over \omega_{2m+1} - \omega_{2n}} = -
\delta_{nk} \; Res_{m=n-1/2} {\pi \cot (m\pi)A_{mn}\over \omega_{2m+1} -
\omega_{2n}}.
}
The residue is computed to be
\eqn\calceight{\eqalign{
&Res_{m=n-1/2} {\pi\cot(m\pi) A_{mn}\over \omega_{2m+1}
-\omega_{2n}} \cr
&= i \sqrt{2} \pi 
v_n^{II} v_{n-1/2}^I \lim_{m \to n-1/2} { \cot(\pi m) \over 
\omega_{2m+1} - \omega_{2n}} \; Res_{m=n-1/2} {1\over \omega_{2m+1} -
\omega_{2n}} = -{i\pi\over 2\sqrt{2} n}\,, 
}}
and therefore we have verified \tosolveone. 

Finally we may compute $C_{mk}$ using \tosolvethree. We are led to
\eqn\calcnine{\eqalign{
Res_{m=-n-1/2} {A_{mk} \pi \cot(\pi m) \over \omega_{2m+1} + \omega_{2n}}
&= { i \sqrt{2} v_k^{II}\, \pi \over \omega_{2n}+ \omega_{2k}} \; Res_{m=-n-1/2}
{v_m^I \cot(m \pi)\over \omega_{2m+1}+ \omega_{2n}} \cr
&= \left({ i\sqrt{2} v_k^{II} v_n^{II} \over \omega_{2n} + \omega_{2k}}\right)\,
{\pi \omega_{2n}\over 4n}\,.
}}
Thus,
\eqn\calcten{
C_{mk}= 2  \left({v_k^{II} v_m^{II} \over \omega_{2m} +
\omega_{2k}}\right)\,,
}
will ensure \tosolvethree\ as well as having the correct flat
space limit. 

As pointed out in section 3.2, there can be contributions from the
integrals around the branch points $m_{\pm}$, which were depicted as
$K_1$ and $K_2$ in figure 2. 
Thus, we need to show that these integrals do in fact vanish in order
for the contour argument to hold. Consider $A_{mk}$ for example. Let $m
= -1/2 +i\mu/2 + \epsilon e^{i\theta}/2$; this implies that
$\omega_{2m+1} = O(\epsilon^{1/2})$ and thus $A_{mk} = O(\epsilon^{-1/2})$. Therefore,
\eqn\calceleven{
\oint_{K_1,K_2} \, dm \, A_{mk} = O(\epsilon^{1/2}) \,,
}
where the extra factor of $\epsilon$ comes from the integration
measure of course. Hence as $\epsilon \to 0$ we see that the integral
around the branch point does indeed vanish as required. Proofs for the
other sum with $A_{mk}$ and the ones with $B_{mk}$ are entirely analogous.

This completes the proof, that
\Avertex -\Cvertex\ solve the plane-wave open-closed vertex equations
\tosolveone -\tosolvesix.

\subsec{Dirichlet boundary conditions}

Using the above line of thought one can also determine the
solution for the vertex with Dirichlet boundary conditions on the
open string. To this effect we introduce the $\mu$-deformed Gamma-function
\eqn\muGammaIcheck{
\check{\Gamma}^{I}_{\mu} (z)
= e^{-\gamma \omega_{2z}/2}\left( {2\over \omega_{2z}+ \mu}\right) 
\prod_{n=1}^\infty \left({\omega_{2n}\over \omega_{2z}+\omega_{2n}} \,
e^{\omega_{2z}/ 2n}\right) \,,
}
which differs from $\Gamma^I_{\mu}(z)$ only in the ``zero-mode''
part. This Gamma-function has again the key property that it satisfies
a reflection identity
\eqn\checkreflec{
\check{\Gamma}^I_{\mu}(z) \check{\Gamma}^I_{\mu}(-z) = -{\alpha\over z \sin(\pi z)} \,,
}
which is proven in the same manner as provided in Appendix B for $\Gamma^I_{\mu}(z)$.
Further we introduce generalizations of the functions $u_m$
\eqn\vchecks{
\check{v}_m^I = {(2m+1) \over \omega_{2m+1}} \  
{\check{\Gamma}^I (m+1/2) \over \sqrt{\pi}\; \Gamma^{II}(m+1)}\,,\qquad
\check{v}_m^{II} = {2\over \omega_{2m}}\  { \Gamma^{II}(m+1/2) \over
\sqrt{\pi}\; \check{\Gamma}^I (m)} \,.
}
Note that $\check{v}^I_{-1/2+iy}$ is an odd function of $y$, which in the
Dirichlet case is needed for the integrals along the cuts to vanish. The
functions $\check{v}^{I, II}$ are related to $v^{I, II}$ by a
$\mu$-dependent factor, which tends to 1 as $\mu\to 0$.
The coefficients for the open-closed vertex in this case are
\eqnn\Avertexcheck
\eqnn\Bvertexcheck
\eqnn\Cvertexcheck
$$
\eqalignno{
\check{A}_{mk} &= -i\sqrt{2}  {\check{v}^I_m \check{v}^{II}_k \over (\omega_{2m+1} -
\omega_{2k})}&\Avertexcheck \cr
\check{B}_{mk} &= {\check{v}^I_m \check{v}^I_k \over (\omega_{2m+1} +
\omega_{2k+1})}&\Bvertexcheck \cr 
\check{C}_{mk} &=2 { \check{v}^{II}_m \check{v}^{II}_k \over (\omega_{2m} + \omega_{2k}) }
\,.&\Cvertexcheck 
}
$$
These can be verified in the same manner as in the Neumann case. 
Consider \eg\ the equation for $\check{A}$, \Dtosolveone. The contour method implies
\eqn\checkone{
\sum_{m=0}^{\infty} \check{A}_{mk} { (2m+1)\over \omega_{2m+1} -
\omega_{2n}} = - \delta_{nk} \; Res_{m=n-1/2} {\pi \cot (m\pi)(2m+1)\,  
\check{A}_{mn} \over \omega_{2m+1} -
\omega_{2n}} \,,
}
where the residue is evaluated as
\eqn\checktwo{\eqalign{
&Res_{m=n-1/2} (2m+1) {\pi\cot(m\pi) \check{A}_{mn} \over \omega_{2m+1}
-\omega_{2n}} \cr
&= -i  2 \sqrt{2} n\,  \pi\,  
\check{v}_n^{II} \check{v}_{n-1/2}^I  \lim_{m \to n-1/2} { \cot(\pi m) \over 
\omega_{2m+1} - \omega_{2n}} \; Res_{m=n-1/2} {1\over \omega_{2m+1} -
\omega_{2n}} = i{\pi\over \sqrt{2} }\,, 
}}
and therefore we have verified \Dtosolveone. The other equations
can be proven to hold in a similar manner.


\newsec{Large-$\mu$ asymptotics}

\subsec{Neumann boundary conditions}

In this section we will analyse the large-$\mu$ asymptotics of the
solutions that we have determined. We shall consider $\mu>0$.
One can find the bulk of the details in
Appendix B, where in particular we give the asymptotics of $v^I_m$ and
$v^{II}_m$. Consider equation \Avertex\ now 
\eqn\Asolution{
A_{mk} = i\sqrt{2}  {v^I_m v^{II}_k \over (\omega_{2m+1} -
\omega_{2k})}\,.
}
We find using the asymptotic formulas that
\eqn\Aall{ A_{mk} = { i\sqrt{2} \over \pi } {4 k (\omega_{2m+1}+\omega_1 ) \over
\omega_{2m+1}\omega_{2k}(\omega_{2m+1}-\omega_{2k})(\omega_{2k}+\omega_1) }{\left(\omega_{2k}+\mu\over\omega_{2m+1}+\mu\right)}^{1/2}
{\over} + O \left(
e^{-\mu} \right) \,.
}
From the above we see that the leading asymptotic behaviour of $A_{mk}$ is 
\eqn\Aasym{
A_{mk} \sim {1\over \pi \mu}{4i\sqrt{2} (2k) \over (2m+1)^2-(2k)^2}\,,
}
which agrees with the naive approximation in
\Aasymp. 
For completeness we also give
\eqn\BCasm{\eqalign{
B_{mk} &= {e^{\gamma+2a_1}(\omega_{2m+1}+\omega_1)(\omega_{2k+1}+\omega_1)
\over \pi\omega_{2m+1}\omega_{2k+1}(\omega_{2m+1}+\omega_{2k+1})
(\omega_{2m+1}+\mu)^{1/2}(\omega_{2k+1}+\mu)^{1/2}} 
+ O\left(e^{-\mu} \right) 
\,,\cr
C_{mk} &= 2 { e^{-\gamma -2a_1} (16 mk) (\omega_{2m}+\mu)^{1/2}
(\omega_{2k}+\mu)^{1/2} \over \pi \omega_{2m} \omega_{2k}(\omega_{2m}+\omega_{2k}) (\omega_{2m}
+\omega_1)(\omega_{2k}+\omega_1) } + O \left(
e^{-\mu} \right)   \,.
}}
We deduce that
\eqn\Basm{
B_{mk} \sim 
{e^{\gamma+2a_1} \over \pi \mu^2}  
\,,
}
as well as
\eqn\Casm{ 
C_{mk} \sim { 8 mk e^{-\gamma -2a_1} \over \pi \mu^4}  \,.
}
For the reader's convenience we give $A_{mk}$ to $O(1 / \mu^5)$ more
explicitly,
\eqn\Aexplasm{ 
A_{mk} =
{8 i \sqrt{2} k\over \pi (- (2k)^2 + (2m+1)^2)}
\left(  {1\over \mu} -
{ (12 k^2 + (2m+1)^2)\over 8 \mu^3} \right)
+ O \left({1\over \mu^5}\right)
\,.
}
As we have emphasised the naive approximation leads one to conclude
that $B,C$ behave as $O(1/\mu^3)$, whereas the analysis above
disagrees with this. One can think of this discrepancy as arising from
a renormalisation due to the higher modes \refs{\klebanovspradlinvolovich}.
We have thus determined the Neumann matrices $A,B,C$ up to
$O(e^{-\mu})$. One should note the similarity of these expression to the ones for the large $\mu$ Neumann matrices in \refs{\he}.

\subsec{Dirichlet boundary conditions}

Similarly, one obtains the asymptotics of the Neumann coefficients in
the case of Dirichlet boundary conditions. They are given by
\eqn\ABCcheckasym{\eqalign{
\check{A}_{mk} &= - { i2 \sqrt{2} \over \pi } { (2m+1) (\omega_{2m+1}+\omega_1)
(\omega_{2k}+\mu)^{3/2} \over \omega_{2m+1} \omega_{2k}
(\omega_{2k}+\omega_1) (\omega_{2m+1}+\mu)^{3/2}
(\omega_{2m+1}-\omega_{2k}) } + O \left(
e^{-\mu} \right) \cr
\check{B}_{mk} &= { e^{\gamma+2a_1}
(2m+1)(2k+1)(\omega_{2m+1}+\omega_1)(\omega_{2k}+\omega_1) \over \pi
\omega_{2m+1} \omega_{2k+1} (\omega_{2m+1}+\mu)^{3/2}
(\omega_{2k+1}+\mu)^{3/2} (\omega_{2m+1}+\omega_{2k+1})} + O \left(
e^{-\mu} \right) 
\,,\cr
\check{C}_{mk} &= { 8e^{-\gamma -2a_1} (\omega_{2m}+\mu)^{3/2}
(\omega_{2k}+\mu)^{3/2} \over \pi \omega_{2m} \omega_{2k} (\omega_{2m}
+\omega_1)(\omega_{2k}+\omega_1) (\omega_{2m}+\omega_{2k})} + O \left(
e^{-\mu} \right)   \,.
}}
In particular, the first terms in the asymptotic expansion are thus
\eqn\Bcheckasm{
\check{B}_{mk} \sim {e^{\gamma+2a_1}(2m+1)(2k+1) \over 4\pi \mu^4}
\,,}
as well as
\eqn\Ccheckasm{ 
\check{C}_{mk} \sim { 8 e^{-\gamma -2a_1} \over \pi \mu^2} 
\,,}
and $\check{A}_{mk}$ to $O(1 / \mu^5)$ is given by
\eqn\Acheckexplasm{ 
\check{A}_{mk} =- {4i\sqrt{2} (2m+1) \over \pi ((2m+1)^2-(2k)^2)} \left(
{1\over \mu} - {3(2m+1)^2 + (2k)^2\over 8\mu^3} + O \left(
{1\over \mu^5} \right) \right)\,.}
The first order term in this expansion is again in agreement with the
gauge theory analysis in \refs{\gomis}. However, the asymptotics for
$\check{B}$ and $\check{C}$ differ again from the naive approximation.


\newsec{Discussion}

In this paper, we constructed the solutions for the Neumann matrices
of the open-closed vertex in the plane-wave light-cone string field
theory for all values of $\mu$.
Complex analytic methods were invoked in order to derive these
solutions and along the way, we were led to define a set of new,
$\mu$-deformed Gamma functions.
In summary, the exact bosonic
vertex for Neumann boundary conditions was shown to be
\eqn\Vfinal{
|V\rangle=\exp(\Delta_1+\Delta_2)|\Omega\rangle\,,
}

\fig{Graph of $\check{A}_{mk}(\mu)$.}{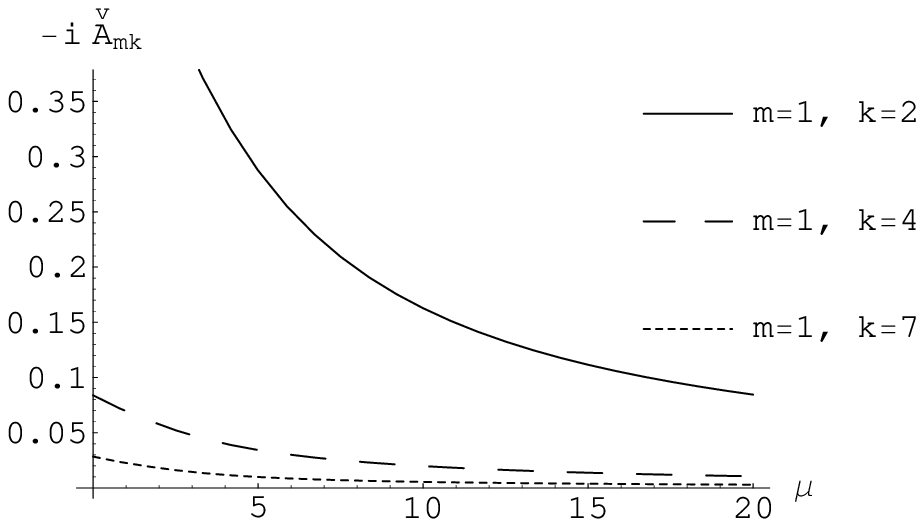}{4.2truein}
\fig{Graph of $\check{B}_{mk}(\mu)$.}{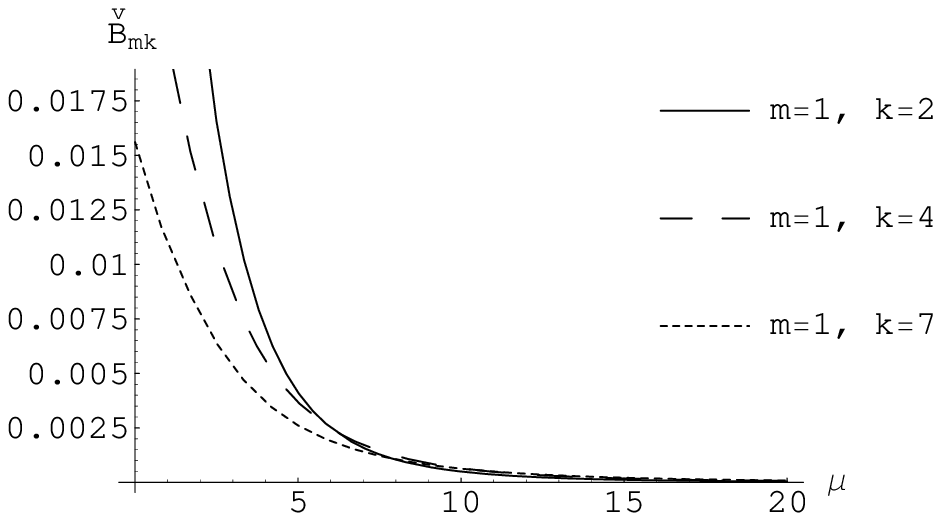}{4.2truein}
\fig{Graph of $\check{C}_{mk}(\mu)$.}{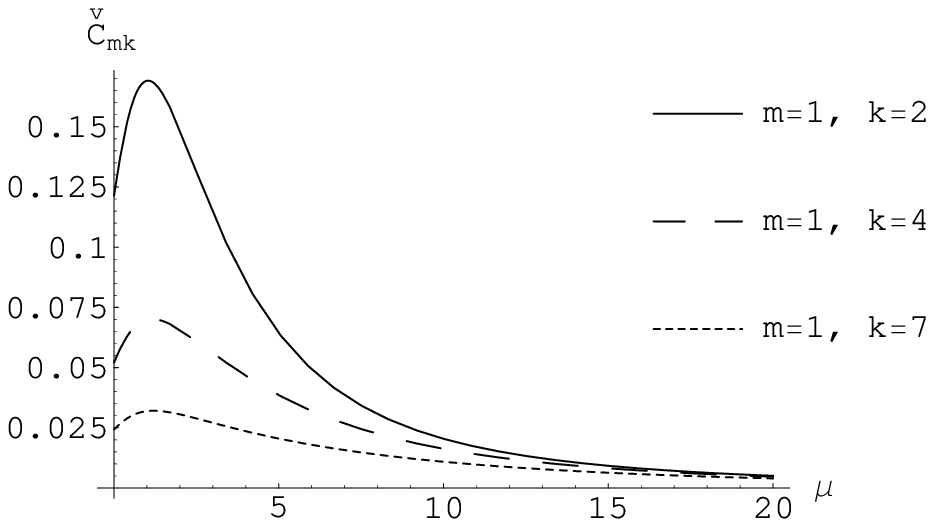}{4.2truein}

\noindent
where
\eqn\Donefinal{
\Delta_1 = -\sum_{m=1}^\infty {\sqrt{2}\over \omega_{2m}} \beta_{-2m}
\alpha^I_{-m}\,,
}
and 
\eqn\Dtwofinal{
\Delta_2 = \sum_{m,n} A_{mn} \beta_{-2m-1} \alpha^{II}_{-n} + \half B_{mn} \beta_{-2m-1}\beta_{-2n-1} + \half C_{mn} \alpha_{-m}^{II} \alpha_{-n}^{II} \,.
}
The Neumann matrices $A,B,C$ were determined to be 
\eqn\Neufinal{\eqalign{
A_{mk} &= i\sqrt{2}  {v^I_m v^{II}_k \over (\omega_{2m+1} -
\omega_{2k})} \cr
B_{mk} &= {v^I_m v^I_k \over (\omega_{2m+1} +
\omega_{2k+1})}\cr 
C_{mk} &= 2 { v^{II}_m v^{II}_k \over (\omega_{2m} + \omega_{2k}) }
\,,
}}
where $v^{I}_m$ and $v^{II}_m$ are the $\mu$-deformed generalizations of
the corresponding functions in flat space, $u_m$. A similar expression
for the Dirichlet case was obtained. In contrast to flat space, the Neumann and Dirichlet solutions differ. It
would be interesting to understand this from the point of view of
T-duality in plane-wave backgrounds \refs{\Michelson,\Japanese}.

Figures 3, 4 and 5 show the behaviour of the Neumann coefficients for Dirichlet boundary conditions,
$\check{A}_{mk}, \check{B}_{mk}$ and $\check{C}_{mk}$, for fixed $m,k$ as a function of $\mu$. 
It is clear from the behaviour of $\check{C}_{mk}$ that there appears to be a
maximum for a particular value of $\mu$, the physical significance of
which still needs to be elucidated. It is also clear from the
asymptotic behaviour of the graphs that $\check{A}$ falls off the slowest
followed by $\check{C}$ and then $\check{B}$. This is consistent with the
large-$\mu$ asymptotics analysed in the previous section.

In the light of the BMN correspondence, we discussed the large-$\mu$
asymptotics of the Neumann matrices and found the results for $A$ and $\check{A}$ to agree in both
the Dirichlet and Neumann case with
\refs{\gomis}, where the corresponding
result was computed in the gauge theory. Note however that our results for
$B$ and $C$ in both Neumann and Dirichlet case differ from the naive approximation.
It would be very interesting to check the next to leading
order corrections on the gauge theory side to confirm the accuracy of our
results.

Using the exact Neumann matrices for the bosonic part of the vertex,
it should now be straightforward to determine the prefactor exactly and
extend our results to the complete superstring vertex. It would be
interesting to study the $\mu$-dependence of the scattering
amplitudes that can be exactly computed.

Finally, in view of the analytic methods applied in this paper, it may
be possible to simplify the derivation of the 
exact Neumann coefficients for the cubic vertex \refs{\he}. 

\vskip1cm

\centerline{{\bf Acknowledgments}}
\pano
We thank Michael Green and Marc Spradlin for useful comments on the
manuscript. We would also like to thank Hari Kunduri and Jasbir Nagi
for comments on an earlier draft. JL is funded by EPSRC. SSN thanks
DAMTP for hospitality. AS is supported by the Gates
Cambridge scholarship and the Perse scholarship of Gonville and Caius College, Cambridge.


\appendix{A}{Collection of useful formulae}

In this appendix we collect some useful formulae ({\it cf.}
\refs{\GR}, \refs{\GSstringfield}).
Stirling approximation
\eqn\stirlingapp{
\Gamma(z)\quad  \sim^{^{\hskip-10pt z\rightarrow\infty \hskip2pt}}
 \sqrt{2\pi}\, z^{z-1/2}e^{-z}\,, \qquad |{\rm arg}z| < \pi \,.} 
The reflection identity is given by
\eqn\reflectnormal{
\Gamma (z) \Gamma (1-z) = {\pi\over \sin (\pi z)}\,.
}
Some useful identities relevant to the flat space section are collected
here.
Defining
\eqn\umflat{
u_n={\Gamma(n+{1\over 2})\over \sqrt{\pi} \Gamma(n+1)}\,,
}
it can be shown that the $u_n$'s satisfy the following identity which
can be readily verified using the contour integration technique
explained in the paper.
\eqn\idone{
\sum_{m=0}^{\infty}{u_m\over m+a+1}={\sqrt{\pi}\Gamma(a+1)\over
\Gamma(a+{3\over 2})}\,.
}
Using this it can be shown that
\eqn\ids{\eqalign{
\sum_{m=0}^{\infty} { u_m\over m-n+\half}&=0, \quad n=1,2,\cdots \cr
\sum_{m=0}^{\infty}{u_m u_n\over m+n+1}&={1\over n+\half}, \quad
n=0,1,2,\cdots \cr
\sum_{m=0}^{\infty}{u_m\over m+n+\half}&= \pi u_n, \quad
n=0,1,2,\cdots \cr
\sum_{m=0}^{\infty}{ u_m u_p\over (m-n+\half)(m-p+\half)}&=
{\pi\over n}\delta_{np}, \quad n,p=1,2,\cdots \,.
}}

\subsec{Sample application of the contour method}

As a warm-up and to illustrate the contour method of section 3, we shall provide a proof of (D.3) in \refs{\GSstringfield} 
\eqn\dthreegs{
\sum_{m=0}^\infty {u_m \over m-n+1/2} =0 \,,
}
using the above method. Consider thus the following integral
\eqn\contint{
\oint_{{C}_1}\,  d q \, \cos (\pi q)\,  \Gamma (-q) \Gamma (q+1/2) \, {1\over q-n+1/2}  \,,
}
which has residues at $q\in \Nop$ and $q= n-1/2$. The contour is
depicted in figure 6.
\fig{Contour $C_1$.}{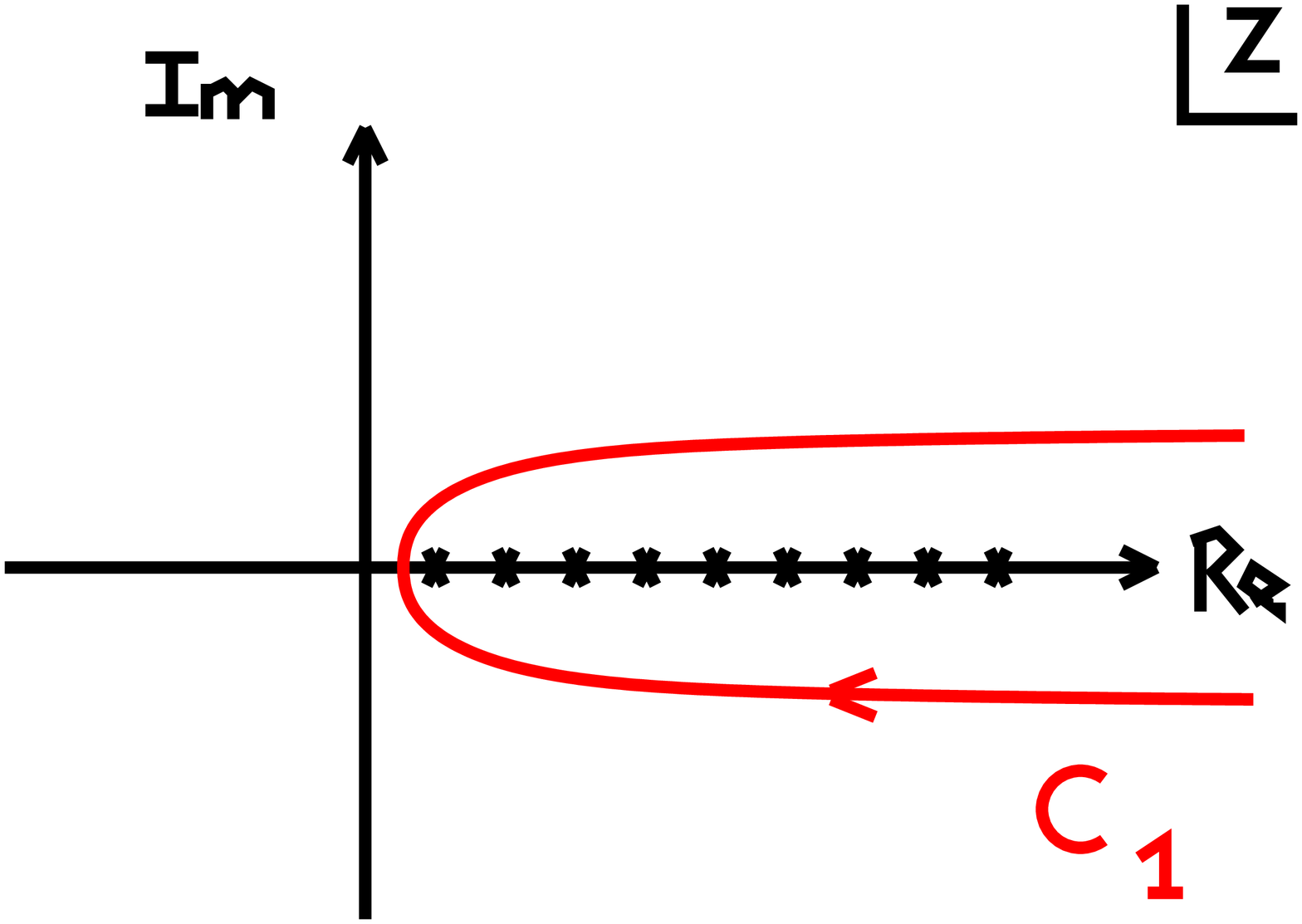}{2.2truein}

The first type of residues, \ie, $q\in \Nop$, will reproduce the sum in \dthreegs. The
latter residue vanishes for all $n\in \Nop$. Thus we are left with the
evaluation of the contour integral in order to show
\dthreegs. Using Stirling's formula, the integrand behaves like
$q^{-3/2}$ for large modulus of $q$. Thus, the contour can be deformed
to infinity, \ie, ${\cal C}_R$ and this integral can be
shown to vanish. This completes the proof.


\appendix{B}{$\mu$-deformed Gamma functions}

\subsec{Definitions and identities}

Here we define two functions, each of which reduce to the Gamma function as $\mu
\to 0$. First recall that the standard Gamma function may be defined
by its Weierstrass product
\eqn\weierstrass{
{1\over \Gamma(z)} = ze^{\gamma z} \prod\limits_{n=1}^{\infty} \left(1+{z\over n}\right)
e^{-z/n} \,,
}
where $\gamma$ is the Euler constant. Now let us define the following
$\mu$-deformed Gamma-function
\eqn\jlone{
{1\over \Gamma^I_{\mu}(z)} = z e^{\gamma
\omega_{2z} /2} \prod\limits_{n=1}^{\infty} \left( {\omega_{2n} +
\omega_{2z}\over \omega_{2n}} \right) e^{-\omega_{2z}/ 2n} \,.}
We will define this to be the {\it $\mu$-deformed Gamma function of the first
kind}. Note that $\omega_z = \sqrt{z^2 +\mu^2}$, where we choose the
finite branch cut for the square root so that $\omega_{-z}=-\omega_z$.
This implies $\omega_n = { \rm sgn}(n) \sqrt{n^2 +\mu^2}$, for $n \in \Zop$. In order to show that $\Gamma^{I}_{\mu}(z)$ is
indeed well-defined, one can use similar arguments as those for
$\Gamma(z)$ (see \refs{\WW} p. 235-236). For
completeness we give an argument. Observe that the factors in the
infinite product go as $1+O(1/n^2)$ for sufficiently large $n$; this
implies the infinite product converges absolutely and uniformly
\foot{Recall that $\prod\limits_n
(1+a_n)$ converges absolutely iff $\sum_n a_n$ converges
absolutely.}.
Notice that $\Gamma^{I}_{\mu}(z)$
has simple poles at $z=-1,-2,\cdots$,  and branch points at $z=\pm i\mu /2$
and a branch cut on $[i\mu /2 , -i\mu /2]$.

Next we will derive a generalisation of the reflection identity
\reflectnormal. Note that
\eqn\jltwo{\eqalign{
&{1\over \Gamma^{I}_{\mu}(z) \Gamma^{I}_{\mu}(-z)} \cr
&= z e^{\gamma \omega_{2z} /2} \prod\limits_{n=1}^{\infty} \left(1+{\omega_{2z}\over \omega_{2n}}\right)
e^{-\omega_{2z}/ 2n} (-z)\, e^{-\gamma \omega_{2z}/2} \prod\limits_{n=1}^{\infty} \left(1-{\omega_{2z}\over \omega_{2n}}\right)
e^{\omega_{2z}/ 2n}  \cr
&= -z^2 \prod\limits_{n=1}^{\infty} \left( { (2n)^2 - (2z)^2\over (2n)^2 +\mu^2} \right)
= -z \sin(\pi z)  {\mu\over 2\sinh(\pi \mu /2)}\,,
}}
where the nice formula $\sin(\pi z) = \pi z
\prod\limits_{n=1}^{\infty}(1-{z^2\over n^2})$ has been used. Thus the
reflection identity takes the form
\eqn\jlthree{
\Gamma^I_{\mu}(z) \Gamma^I_{\mu}(-z) = -{\alpha\over z\sin(\pi z)}\,,
}
where we have defined $\alpha = 2\sinh (\pi\mu /2) / \mu$.
It is easy to see that it reduces to the usual identity when $\mu \to
0$. One
can use this identity to calculate the residues of this new
function. For $n \neq 0$
\eqn\jlfour{\eqalign{
Res_{z=-n} \Gamma^{I}_{\mu}(z) &= \lim_{z \to -n}
(z+n)\Gamma^{I}_{\mu}(z)  \cr
&= \lim_{z \to -n} {\alpha (z+n)\over -z \sin(\pi z) \Gamma^{I}_{\mu}(-z)} \cr
&= { \alpha \over \pi }
{(-1)^n\over n \Gamma^{I}_{\mu}(n)}\,.
}}
Finally, one more crucial property of this function needs to be
mentioned. $\Gamma^{I}_{\mu}( iy)$ is an odd function of $y$, on
both sides of the branch cut, albeit not the same function on either
side! 

For the solution of the plane-wave open-closed vertex, we shall need
 an additional kind of $\mu$-deformed Gamma-function.  Define the { \it $\mu$-deformed Gamma
 function of the second kind} as follows
\eqn\jlfive{
{1\over \Gamma^{II}_{\mu}(z)} = \left( { \omega_{2z-1} +
\omega_1\over 2} \right)
 e^{\gamma (\omega_{2z-1} + 1)/2} \prod\limits_{n=1}^{\infty} \left(
{\omega_{2z-1}+\omega_{2n+1}\over \omega_{2n}} \right)
e^{-(\omega_{2z-1}+1)/2n}\,,
}
which obviously satisfies $\Gamma^{II}_{\mu=0}(z) = \Gamma(z)$. Note
that in this definition it is not obvious that the infinite product
converges. However we can
prove that it does, although convergence relies crucially on the
exponential factors. The argument runs as follows; for large $n$
\eqn\jlsix{\eqalign{
&\left(
{\omega_{2z-1}+\omega_{2n+1}\over \omega_{2n}} \right)
e^{-(\omega_{2z-1}+1)/ 2n} = {\omega_{2n+1}\over \omega_{2n}}
\left( 1+ {\omega_{2z-1}\over \omega_{2n+1}} \right) \left( 1- {
\omega_{2z-1} + 1\over 2n} + O \left({1\over n^2} \right)
\right) \cr
&= \left( 1 + {1\over 2n} \right) \left( 1+ O \left({1\over n^2}
\right) \right) \left( 1 + { \omega_{2z-1} \over 2n+1} + O \left({1\over n^2}
\right) \right) \left( 1- {
\omega_{2z-1} + 1\over 2n} + O \left({1\over n^2} \right) 
\right) \cr
&= \left( 1 + {1\over 2n} \right) \left( 1- {1\over 2n} -
{\omega_{2z-1}\over 2n(2n+1)} +  O \left({1\over n^2} \right)
\right) = 1 + O \left( {1\over n^2} \right)\,,
}}
and hence the infinite product converges absolutely and uniformly. The function
 $\Gamma^{II}_{\mu}(z)$ has simple poles at $z=0,-1,-2, \cdots$, branch
 points at $1/2 \pm i\mu /2$ and a branch cut on $[1/2-i\mu/2,1/2+i\mu/2]$.
 Interestingly a reflection identity exists for this function as well
\eqn\Gtworeflect{
\Gamma^{II}_{\mu} (1+z) \Gamma^{II}_{\mu}(-z) =
-{\alpha \over \sin (\pi z)}\,.
}
Finally we should emphasise
that $\Gamma^{II}_{\mu}(1/2 +iy)$ is even in $y$ on both sides of the
branch cut.

For the solution to the vertex with Dirichlet boundary conditions we
further introduce a $\mu$-deformed Gamma function
\eqn\Gammacheckapp{
{1\over \check{\Gamma}^{I}_{\mu} (z)}
=  \left({\omega_{2z}+ \mu\over 2} \right)e^{\gamma
\omega_{2z} /2} \prod\limits_{n=1}^{\infty} \left( {\omega_{2n} +
\omega_{2z}\over \omega_{2n}} \right) e^{-\omega_{2z}/ 2n} \,,
}
which differs from $\Gamma^I_\mu(z)$ only in the ``zero-mode''
term. It is straightforward to prove that this satisfies the same
reflection identitiy as $\Gamma^I_\mu(z)$. Note however, that
$\check{\Gamma}^{I}_{\mu}( iy)$ is an even function of $y$ on both
sides of the branch cut. This is the key difference to $\Gamma^I_\mu$,
and ensures that in the derivation of the Neumann coefficients, the
integrals along each of the branch cuts vanish for the case of
Dirichlet boundary conditions.

Note that generalisations of the identity $\Gamma(z+1) = z\Gamma(z)$
have not been found for any of the $\mu$-deformed Gamma-functions.

\subsec{Large-$\mu$ asymptotics}

In this section we will develop the necessary tools to calculate the
large-$\mu$ asymptotics of certain combinations of our new
functions. Namely we are interested in the behaviour of $v^I_m$ and
$v^{II}_m$, which are defined as
\eqn\vonem{\eqalign{
v_m^I &= {(2m+1) \over \omega_{2m+1}} \  
{\Gamma^I (m+1/2) \over \sqrt{\pi}\; \Gamma^{II}(m+1)} \cr
v_m^{II} &= {2\over \omega_{2m}}\  { \Gamma^{II}(m+1/2) \over \sqrt{\pi}\;
\Gamma^I (m)}\,.
}}
Using the definitions of our Gamma functions we can be more explicit
\eqn\vexplicit{\eqalign{
v_m^I &=  {e^{\gamma/2} \over \sqrt{\pi}} {\omega_{2m+1} + \omega_1\over  \omega_{2m+1}}
\prod_{n=1}^{\infty} \left( {\omega_{2m+1} +
\omega_{2n+1}\over \omega_{2m+1} + \omega_{2n}} \right) e^{-1/ 2n}
\cr
v^{II}_m &= {2\over \omega_{2m}} {e^{-\gamma
/2}\over \sqrt{\pi}} {2m \over \omega_{2m} + \omega_1} \prod_{n=1}^{\infty} \left( {\omega_{2m} + \omega_{2n}\over \omega_{2m} +
\omega_{2n+1}} \right) e^{1/ 2n} \,,
}}
and we see that it is sufficient to study the following infinite
product
\eqn\Sz{
e^{S_z} \equiv \prod_{n=1}^{\infty} \left( {\omega_{z} +
\omega_{2n+1}\over \omega_{z} + \omega_{2n}} \right) e^{-1/ 2n}\,.
}
Taking the logarithmic derivative of this infinite product with
respect to $\mu$ we get
\eqn\Szasymp{
{\partial S_z\over \partial \mu} = {\mu\over \omega_{z}}\sum_{n=1}^{\infty} {1\over \omega_{2n+1}} -
{1\over \omega_{2n}}\,.
} Let
\eqn\Rdef{
R=\sum_{n=1}^{\infty} {1\over \omega_{2n+1}} -
{1\over \omega_{2n}} \,,
}
and differentiating with respect to $\mu$, we get
\eqn\Rexp{
{\partial R\over \partial \mu} = \sum_{n=1}^{\infty} -{\mu\over \omega_{2n+1}^3} +
{\mu\over \omega_{2n}^3} \,,
}
and thus we see that all we need now is the asymptotic behaviour of 
\eqn\sumcubed{\sum_{n=1}^{\infty} {1\over \omega_{2n}^3} \quad, \quad \sum_{n=1}^{\infty} {1\over \omega_{2n+1}^3}\,,
}
which can be worked out as follows. One may obtain an integral
representation of the first sum using the integral definition of $\Gamma(3/2)$, \ie, making use of 
\eqn\jlten{
\Gamma (z) = x^z \int\limits_0^\infty e^{-xt} t^{z-1} dt \,,
}
which holds for Re$(z)>0$ and Re$(x)>0$. Thus we can write
\eqn\jleleven{
\sum_{n=1}^{\infty} {1\over \omega_{2n}^3} = {2\over \sqrt{\pi}}
\int_0^{\infty} dt \; t^{1/2} e^{-\mu^2 t} \sum_{n=1}^{\infty} e^{-4n^2 t}\,,
}
and we see that we have something related to a theta function in the
integrand. In fact we have $\sum_{n=1}^{\infty} e^{-4n^2 t} =
(\psi(4t/\pi)-1)/2$ where $\psi(t) = \sum_{n=-\infty}^{\infty}
e^{-n^2\pi t}$, which has the nice transformation law $\psi(t)=
{1\over \sqrt{t}} \psi(1/t)$. Now if we change variables to $s =
\mu^2 t$ it is easy to deduce
\eqn\jltwelve{\eqalign{
\sum_{n=1}^{\infty} {1\over \omega_{2n}^3} &= - {1\over 2\mu^3}+ {1\over \mu^3 \sqrt{\pi}}
\int_0^{\infty} ds \; s^{1/2} e^{- s} \psi(4s/(\pi \mu^2)) \cr
&= - {1\over 2\mu^3} + {1\over 2\mu^2} \int_0^{\infty} ds \; e^{-s}
\psi(\mu^2 \pi / 4s)\,,
}}
where we have used the transformation law for $\psi(t)$ in the second equality.
Now since $\lim_{t \to \infty} \psi(t) =1$ we deduce that
\eqn\omegacubed{
\sum_{n=1}^{\infty} {1\over \omega_{2n}^3} \sim {1\over 2\mu^2}\,,
}
for $\mu \to \infty$.
In fact we have a much stronger result.
This is easily derived from our integral representation as follows. We
have\foot{Related asymptotics have been discussed in \refs{\he}.}
\eqn\evenomega{\eqalign{
\sum_{n=1}^{\infty} {1\over \omega_{2n}^3} &= - {1\over 2\mu^3} + {1\over 2\mu^2} \int_0^{\infty} ds \; e^{-s}
\psi(\mu^2 \pi / 4s) \cr
&= {1\over 2\mu^2} - {1\over 2\mu^3} + {1\over 2\mu^2}\int_0^{\infty} ds \; e^{-s}
(\psi(\mu^2 \pi / 4s) -1) \cr
&= {1\over 2\mu^2} - {1\over 2\mu^3} + {1\over 2\mu^{2+N}}
\int_0^{\infty} dr \; e^{-r / \mu^N} (\psi(\mu^{2+N} \pi / 4r) -1) \cr
&= {1\over 2\mu^2} - {1\over 2\mu^3} + O \left( {1\over  \mu^{2+N}}
\right) \cr 
&= {1\over 2\mu^2} - {1\over 2\mu^3} + O \left( e^{-\mu} \right) \,,
}}
where we have used the change of variables $r=\mu^N s$ in the third
equality and $N$ can be any positive integer, in particular as large
as we like \foot{Recall that $f(x)=O(g(x))$ if there exists a constant
$C$ such that $|f(x)|< C|g(x)|$ for all $x$ greater than some $x_0$.}.
In a similar manner the
integral representation of $\Gamma(3/2)$ can be used to deal with the
second sum too. We easily show that
\eqn\omone{
\sum_{n=1}^{\infty} {1\over \omega_{2n+1}^3} = -{1\over \mu^3} +
{1\over \sqrt{\pi}} \int_0^{\infty} dt \, t^{1/2} e^{-\mu^2 t}
\theta_2(4it/ \pi)\,,
}
where $\theta_2(\tau) = 2 \sum_0^{\infty} e^{i\pi \tau (n+1/2)^2}$ is
one of the theta functions. Using the modular transformation property
$ \sqrt{-i \tau} \theta_2(\tau) = \theta_4(-1/ \tau)$, it then follows that
\eqn\omtwo{
\sum_{n=1}^{\infty} {1\over \omega_{2n+1}^3} = {1\over 2\mu^2}
-{1\over \mu^3} + {1\over 2\mu^2} \int_0^{\infty} ds \, e^{-s} (
\theta_4(-\mu^2 \pi / (4is)) -1) \,,
} 
where $\theta_4(\tau) = \sum_{-\infty}^{\infty} (-1)^n e^{i\pi \tau
n^2}$. If we make the change of variables in the integral on the RHS
$r = \mu^N s$, it is easy to see that this term is $O(1/ \mu^{N+2})$ for any positive
integer $N$; we have thus derived the useful result
\eqn\omthree{
\sum_{n=1}^{\infty} {1\over \omega_{2n+1}^3} = {1\over 2\mu^2}
-{1\over \mu^3} + O \left( e^{-\mu} \right) \,.
}
Using this together with \evenomega\ allows us to deduce that
\eqn\omfour{
{ \partial S_z \over \partial \mu} = -{1 \over 2\omega_z}+ O \left( e^{-\mu} \right) \,,
}
and therefore
\eqn\omfive{
e^{S_z} = { e^{a_1} \over \sqrt{\mu + \omega_z}} + O \left( e^{-\mu} \right) \,,
}
where $a_1$ is a constant. A priori it seems that $a_1$ could depend on
$z$, however it can be proven it does not. To see this take
the logarithmic derivative with respect to $z$ of \omfive\ and then the limit
$\mu \to \infty$ and compare with the exact expression for ${\partial S_z\over \partial z}$. 
The hard part is now done! We can now straightforwardly find the
expansions for $v^I_m$ and $v^{II}_m$. We get
\eqn\omsix{\eqalign{
v^I_m &= {e^{\gamma/2 +a_1} \over \sqrt{\pi}\, \omega_{2m+1}}
 {\omega_{2m+1}+\omega_1 \over (\omega_{2m+1} + \mu)^{1/2}} + O \left(e^{-\mu} \right) 
 \cr
v^{II}_m &= { e^{-\gamma /2 - a_1} \over\sqrt{\pi}\, \omega_{2m} } {
4m (\omega_{2m}+\mu)^{1/2} \over \omega_{2m} + \omega_1} + O \left( e^{-\mu} \right) \,.
}}
Similarly the expansions for the $\check{v}^I_m$ and
 $\check{v}^{II}_m$ as defined in \vchecks\ are computed to be
\eqn\omsixcheck{\eqalign{
\check{v}^I_m & = {e^{\gamma/2 +a_1}(2m+1) \over \sqrt{\pi}\, \omega_{2m+1}}
 {\omega_{2m+1}+\omega_1 \over (\omega_{2m+1} + \mu)^{3/2}} + O \left(e^{-\mu} \right) 
 \cr
\check{v}^{II}_m & = {2 e^{-\gamma /2 - a_1} \over\sqrt{\pi}\, \omega_{2m} } {
(\omega_{2m}+\mu)^{3/2} \over \omega_{2m} + \omega_1} + O \left( e^{-\mu} \right) \,.
}}

\subsec{More asymptotics}

In this section we provide the asymptotics, which will ensure that the
integrals of the circle at infinity, that arise in the
contour method, do indeed vanish.
Consider $e^{S_z}$. Taking the logarithmic derivative with respect to
$z$ we get
\eqn\omseven{
{\partial S_z\over \partial z} = {z\over \omega_z}
\sum_{n=1}^{\infty} \left( {1\over \omega_z + \omega_{2n+1} } -
{1\over \omega_z + \omega_{2n} } \right)\,.
}
Let us define
\eqn\omeight{
K_z =  \sum_{n=1}^{\infty} \left( {1\over \omega_z + \omega_{2n+1} } -
{1\over \omega_z + \omega_{2n} } \right)\,,
}
so that
\eqn\omnine{
{\partial K_z\over \partial z} = {z\over \omega_z} \sum_{n=1}^{\infty}
{1\over (\omega_z + \omega_{2n})^2} - {1\over (\omega_z +
\omega_{2n+1})^2}\,.
}
We will now find an integral representation for each of the sums on
the RHS. We will denote
\eqn\Sonetwo{\eqalign{
S_1(a) &= \sum_{n=1}^{\infty} {1\over (a + \omega_{2n})^2}\,, \cr
S_2(a) &= \sum_{n=1}^{\infty} {1\over (a + \omega_{2n+1})^2}\,.
}}
First consider evaluating $S_1(a)$, for which we will again use complex
methods. We have (if Re$(a) >0$ \foot{For Re$(a)<0$ we
close in the left hand side of the plane and run along the left hand
side of the branch cut. This will give the same answer.})
\eqn\Soneasym{
S_1(a) = - \oint_C {dz\over 2 \pi i} \, {\pi \cot(\pi z)\over (a+
\omega_{2z})^2} \,,
}
where $C$ is the contour depicted in figure 7, which runs along the imaginary axis (avoiding
the pole in $\cot(\pi z)$ to the right) just to the right of the finite
branch cut due to the $\omega_{2z}$ and closes to the right in a
semi-circle of radius $R$ enclosing
the whole of the right hand plane as $R \to \infty$. We split the contour up $C$ as
comprising of $C_{\pm}= (\pm i \mu/2, \pm i\infty)$, $C_{\epsilon} = \{
\epsilon e^{i\theta} | \; -\pi /2 \leq \theta \leq \pi /2 \}$ and $C_R 
=  \{ R e^{i\theta} | \; -\pi /2 \leq \theta \leq \pi /2 \}$ and of
course the line integrals along the branch cut (depicted in blue in
figure 4), however these vanish due to the integrand being odd. 

\fig{Contour $C$.}{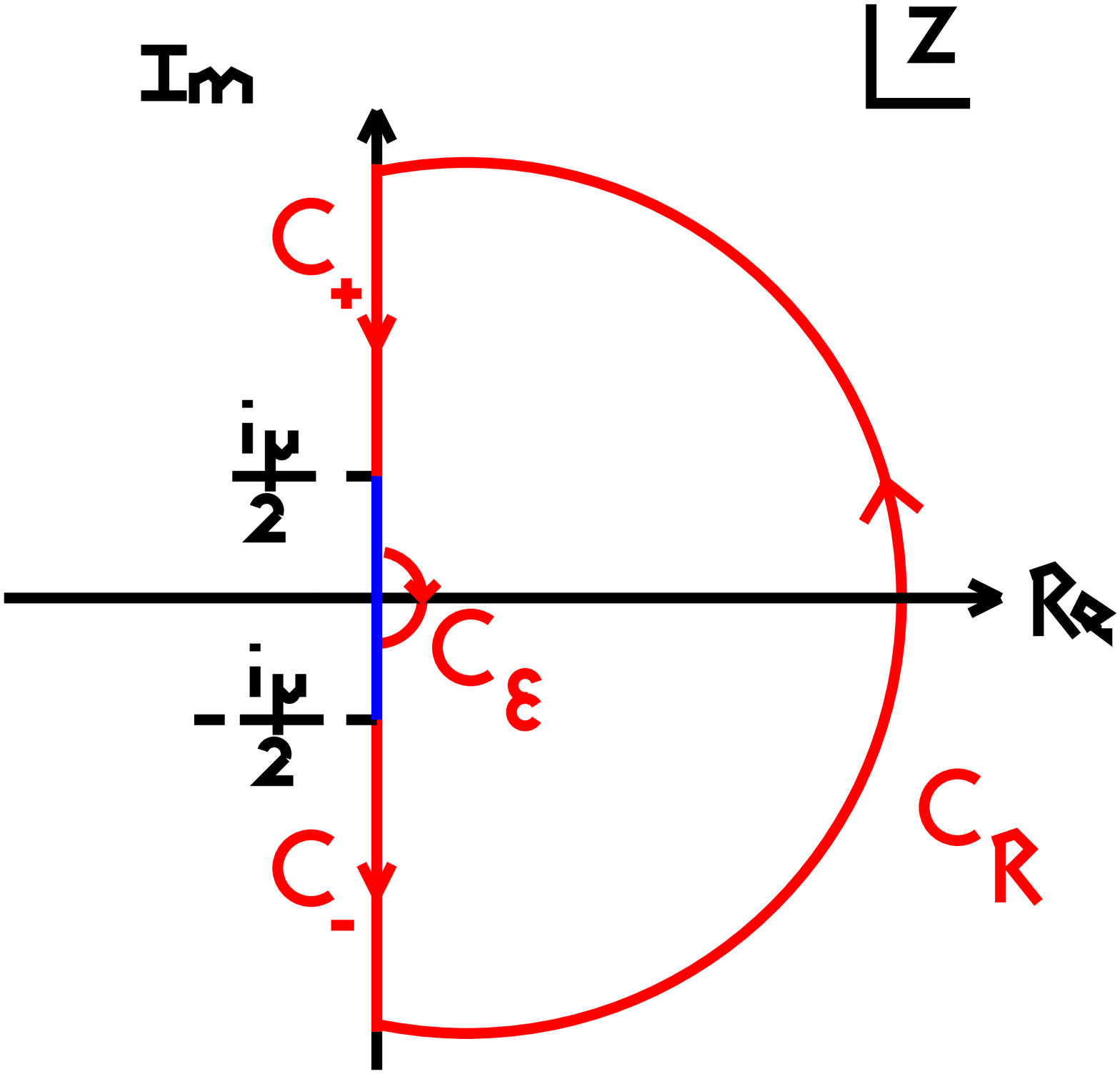}{2.2truein}

Thus the relevant path is $C_+ \cup
C_R \cup
C_- \cup C_{\epsilon}$. We traverse the contour in a clockwise
direction, hence the minus in the integral above. It is easy to see
that the integral along $C_R$ tends to zero as $R \to\infty$. Also we
have
\eqn\doneone{
\lim_{\epsilon \to 0} \oint_{C_{\epsilon}} {dz\over 2 \pi i} \,
{\pi \cot(\pi z)\over (a+\omega_{2z})^2} = {1\over 2(a+\mu)^2}\,.
}
Finally noting that $\omega_{2z}$ is $\sqrt{\mu^2 -4y^2}$ along $C_+$
and $-\sqrt{\mu^2 -4y^2}$ along $C_-$, we arrive at,
\eqn\donetwo{
\oint_{C_+ \cup C_-}{dz\over 2 \pi i} \,
{\pi \cot(\pi z)\over (a+\omega_{2z})^2} = -2a \int_{\mu /2}^{\infty}
dy \; { \coth (\pi y) \sqrt{4y^2 -\mu^2}\over (a^2 - \mu^2 +4y^2)^2}.
}
All this means is that
\eqn\donethree{
S_1(a)= - {1\over 2(a+\mu)^2} + 2a \int_{\mu /2}^{\infty}
dy \; { \coth (\pi y) \sqrt{4y^2 -\mu^2}\over (a^2 - \mu^2 +4y^2)^2}.
}
By a completely analogous method one may show that
\eqn\donefive{
S_2(a) = -{1\over (a+\omega_1)^2} + 2a \int_{\mu /2 }^{\infty} dy \,
{ \tanh(\pi y)\sqrt{4y^2 -\mu^2}\over (a^2 -\mu^2 +4y^2)^2}\,.
}
By writing $\coth(\pi y) = 1+ 2/(e^{2\pi y} -1)$ and $\tanh(\pi y) = 1-
2/(e^{2 \pi y} +1)$ one may convince oneself that the remaining integrals
in our expression for $S_1(a)-S_2(a)$ are $O(1/a^3)$ and hence that
\eqn\donesix{
S_2(a)-S_1(a)= -{1\over 2a^2} + O \left( {1\over a^3} \right)\,.
}
Therefore we get
\eqn\doneseven{
K_z = -{1\over 2z} + O\left( {1\over z^2} \right)\,,
}
since the integration constant must be zero. Finally we have
\eqn\doneeight{
S_z = -{1\over 2}\log z + c_3 + O \left( {1\over z} \right)\,,
}
thus proving that $v^I_m$ and $v^{II}_m$ have the same asymptotics as
$u_m$ which is of course crucial in order to justify the contour method.


\listrefs

\bye